\documentclass[a4paper,fleqn,usenatbib]{mnras}

\usepackage[T1]{fontenc}
\usepackage{ae,aecompl}

\usepackage{graphicx}	
\usepackage{amsmath}	
\usepackage{amssymb}	
\usepackage{times}
\usepackage{orcidlink}
\usepackage{newtxtext,newtxmath}

\defcitealias{Ward2020_HI}{Paper I}

\title[Molecular cloud lifetimes in the LMC]{Towards a multi-tracer timeline of star formation in the LMC -- II. The formation and destruction of molecular clouds}

\author[J.~L.~Ward et al.]{Jacob L.~Ward,$^{1}$\thanks{E-mail: jakelward1@gmail.com (JLW)}
J.~M.~Diederik~Kruijssen\orcidlink{0000-0002-8804-0212},$^1$
M\'{e}lanie~Chevance\orcidlink{0000-0002-5635-5180},$^{1,2}$
Jaeyeon~Kim\orcidlink{0000-0002-0432-6847}$^1$
\newauthor
and Steven~N.~Longmore$^3$
\\
$^{1}$Astronomisches Rechen-Institut, Zentrum f\"{u}r Astronomie der Universit\"{a}t Heidelberg, M\"{o}nchhofstra{\ss}e 12-14, 69120 Heidelberg, Germany\\
$^{2}$Instit\"ut f\"{u}r Theoretische Astrophysik, Zentrum f\"{u}r Astronomie der Universit\"{a}t Heidelberg, Albert-Ueberle-Stra{\ss}e 2, 69120 Heidelberg, Germany\\ 
$^3$Astrophysics Research Institute, Liverpool John Moores University, IC2, Liverpool Science Park, 146 Brownlow Hill, Liverpool L3 5RF, UK
}

\date{Accepted 2022 August 15. Received 2022 August 15; in original form 2022 June 17}

\pubyear{2022}

\begin{document}
\label{firstpage}
\pagerange{\pageref{firstpage}--\pageref{lastpage}}
\maketitle

\begin{abstract}
The time-scales associated with various stages of the star formation process represent major unknowns in our understanding of galactic evolution, as well as of star and planet formation. This is the second paper in a series aiming to establish a multi-tracer time-line of star formation in the Large Magellanic Cloud (LMC), focusing on the lifecycle of molecular clouds. We use a statistical method to determine a molecular cloud lifetime in the LMC of $t_{\text{CO}}=11.8^{+2.7}_{-2.2}$~Myr. This short time-scale is similar to the cloud dynamical time, and suggests that molecular clouds in the LMC are largely decoupled from the effects of galactic dynamics and have lifetimes set by internal processes. This provides a clear contrast to atomic clouds in the LMC, of which the lifetimes are correlated with galactic dynamical time-scales. We additionally derive the time-scale for which molecular clouds and H\,{\sc ii} regions co-exist as $t_{\text{fb}}=1.2^{+0.3}_{-0.2}$~Myr, implying an average feedback front expansion velocity of 12\,km\,s$^{-1}$, consistent with expansion velocities of H\,{\sc ii} regions in the LMC observed directly using optical spectroscopy. Taken together, these results imply that the molecular cloud lifecycle in the LMC proceeds rapidly and is regulated by internal dynamics and stellar feedback. We conclude by discussing our measurements in the context of previous work in the literature, which reported considerably longer lifetimes for molecular clouds in the LMC, and find that these previous findings resulted from a subjective choice in timeline calibration that is avoided by our statistical methodology.
\end{abstract}




\section{Introduction}

The molecular cloud lifecycle is the key process through which stars, clusters and associations form, together driving the evolution of galaxies \citep[e.g.][]{Krumholz2019,Chevance2020b,Chevance2022}. Due to this fundamental role of the cloud-scale baryon cycle in galaxy evolution, the time-scales on which molecular clouds form, evolve, and disperse represent key uncertainties in galaxy evolution models \citep[e.g.][]{Fujimoto2019,Keller2020,Jeffreson2021,Semenov2021}. Estimates of the lifetimes of molecular clouds range from less than a few Myr in the dynamically-dominated regime of the Central Molecular Zone of the Milky Way \citep{Kruijssen2015,Henshaw2016,Barnes2017,Jeffreson2018b} to time-scales based on molecular gas within interarm regions of over 100\,Myr \citep{Scoville1979b,Scoville1979,Koda2009}. Work using cloud-classification frameworks and dynamical arguments place molecular cloud lifetimes in the tens of Myr \citep[e.g.][]{Elmegreen2000,Engargiola2003,Kawamura2009,Meidt2015,Corbelli2017}

Cloud-classification methods are limited by the subjectivity of cloud definitions within the hierarchically-structured interstellar medium (ISM), as well as by the requirement to sufficiently resolve the internal structure of molecular clouds and associated stellar populations. In \citet{KL14}, the \textquoteleft Uncertainty Principle for Star Formation\textquoteright formalism was presented as a method for determining the time-scales associated with the molecular cloud lifecycle that overcomes the limitations of cloud-classification based methods by not relying on a subjective cloud definition and not requiring the internal structure of clouds to be resolved. Instead, it uses the variation of flux ratios as a continuous function of spatial scale to constrain the size and time-scales of the cloud lifecycle (this measurement of flux ratios as a function of spatial scale was pioneered by \citealt{Schruba2010}). The practical application of this new framework is described in detail in \citet{Kruijssen2018}, and it has been used to infer molecular cloud lifetimes in the range 8--30~Myr, exhibiting physical variation with the galactic environment across the nearby galaxy population \citep{Kruijssen2019,Chevance2020,Zabel2020,Kim2021,Kim2022}.

At a distance of 50\,kpc \citep{Laney2012} and with a mass of $5\pm 1\times10^{9}$\,$M_{\sun}$ \citep{Alves2000}, the Large Magellanic Cloud (LMC) is the closest and most massive of the Milky Way's star-forming satellites. It therefore represents an important laboratory for studying the process of star formation from galactic scales down to scales of individual stars and protostars. Consequently, the LMC has been fully sampled at a wider range of wavelengths than any other galaxy in the Universe, making it the first place in which star formation can be studied as a complete process from the condensation of atomic and molecular clouds through to the emergence of main sequence stars and clusters. 

In \citet{Ward2020_HI} (hereafter \citetalias{Ward2020_HI}), we used the statistical methodology of \citet{Kruijssen2018} to infer a lifetime for the atomic H{\sc i} clouds that precede the formation of molecular clouds in the LMC. We obtained a lifetime of $48^{+13}_{-8}$~Myr, consistent with time-scales predicted based on galactic dynamics \citep{Jeffreson2018} and dominated by the time-scale for the gravitational collapse of the mid-plane ISM. In this work, we extend our measurement of the atomic cloud lifetime with a measurement of the molecular cloud lifetime.

Previous estimates of the molecular cloud lifetime in the LMC have been based on the classification of molecular clouds by their stellar content (obtaining lifetimes of 20--30~Myr, \citealt{Kawamura2009}). Our primary goal is to obtain a measurement using the statistical formalism of \citet{Kruijssen2018}, which has now been applied to high-resolution CO and H$\alpha$ maps of more than sixty nearby galaxies. This serves two immediate goals. First, the use of a consistent methodology allows us to combine the resulting molecular cloud lifetime with our prior measurement of the atomic cloud lifetime and that way obtain an end-to-end census of the cloud lifecycle in the LMC, from molecular cloud formation to destruction. Secondly, the comparison of the inferred molecular cloud lifetime to that obtained with the cloud identification/counting method of \citet{Kawamura2009} allows us to determine how to interpret similarities or differences between their cloud lifetimes and those obtained in other galaxies using our new methodology.

The structure of this paper is as follows. In Section~\ref{obsSection}, we outline the observations and methods used to obtain the molecular cloud lifetime. Section~\ref{results} presents our results, summarised as a complete timeline combining the molecular cloud lifetime, condensation time, and feedback time-scale. In Section~\ref{discussion}, we compare the newly derived time-scales with those of previous studies as well as critically assess the robustness of our results. Finally, we summarise our key findings in Section~\ref{conclusions}.

\section{Observations and methodology}

In this section we outline the H$\alpha$, CO, and H{\sc i} observations used in this work. We also briefly introduce the analysis technique that is applied to characterise the molecular cloud lifecycle. 

\subsection{Observations}

\label{obsSection}

Two key CO surveys of the LMC have taken place in recent decades that cover a significant portion of the LMC with sufficient resolution to apply the framework used in this work: the NANTEN CO survey of the LMC \citep{Fukui2008} and the MAGMA survey \citep{TonyWong2011}. 
The NANTEN survey covers approximately 30 deg$^2$ of the LMC. Capitalising on this complete dataset, the MAGMA survey specifically targets sources identified in the NANTEN survey. As a result, a total area of only $\sim$\,3.6\,deg$^2$ is covered, but still recovering 80\% of the total CO flux.
While the NANTEN survey provides better coverage, the data is yet to be made publicly available. Therefore, we are restricted to only using the MAGMA data and cannot perform a direct comparison with the results of the NANTEN survey.\footnote{We investigate the impact of the limited coverage in Appendix~\ref{limitedCoverage}, and find that it does not significantly affect the results of this work.} We use the third MAGMA data release in this study \citep{Wong2017}.

When applying our statistical methodology (see Section~\ref{uncertaintyPrinciple}), we apply a series of masks in order to omit regions where molecular clouds are reported by the NANTEN survey but are not sampled by the MAGMA survey. This is done to maximise the available information by not restricting the analysis to solely the region covered by MAGMA, while simultaneously avoiding the identification of H\,{\sc ii} regions as isolated when they are in fact associated with molecular clouds that fall outside the MAGMA sample. This is addressed in more detail in Section~\ref{uncertaintyPrinciple}. 

As tracers of massive star formation, we make use of two independently constructed H$\alpha$ emission maps of the LMC: the Magellanic Clouds Emission Line Survey (MCELS, \citealt{Smith2005}) H$\alpha$ map and the Southern H-Alpha Sky Survey Atlas (SHASSA, \citealt{Gaustad2001}) H$\alpha$ map. The MCELS emission map uses a narrowband filter centred on 6563\,\AA{} with a FWHM of 30\,\AA{} and is not continuum-subtracted, and therefore contains continuum emission in this wavelength range. The SHASSA H$\alpha$ map uses a narrowband filter also centred on H$\alpha$ with a bandwidth of 32\,\AA{}, but contrary to the MCELS map the SHASSA map is continuum-subtracted. The use of two independently derived H$\alpha$ maps allows for a greater confidence in the inferred time-scales, because the H$\alpha$ emission provides the `reference time-scale' that is used to fix the absolute scale of the inferred evolutionary timeline. Because the MCELS and SHASSA maps include and exclude the continuum, respectively, their reference time-scales differ. In appendix~A of \citetalias{Ward2020_HI}, we demonstrated that the atomic cloud lifetimes independently inferred using both maps are consistent with one another.

Finally, in Section~\ref{mc_formation}, we use the H{\sc i} map of the LMC of \citet{Kim2003} in order to determine the time-scale for which both atomic gas emission and CO emission co-exist. This map was obtained using Parkes and ATCA; for further details we refer to \citetalias{Ward2020_HI}, where we used it to determine the time-scale associated with the atomic gas cloud progenitors of the molecular clouds studied in this work.

\subsection{Application of the uncertainty principle for star formation}

\label{uncertaintyPrinciple}

The \textquoteleft uncertainty principle for star formation\textquoteright\, \citep{KL14,Kruijssen2018} is a fundamental statistical formalism that measures the small-scale variation of the flux ratio between two tracers relative to the large-scale galactic average, and fits a simple forward model to determine the relative lifetimes for which independent units within a galaxy emit in either of the tracer maps, as well as the relative time for which these tracers may coexist, and the mean separation length of these independent units. The formalism avoids the issues of stringent resolution requirements and subjective classification methods that affected previous methods, by utilising the variation of the gas-to-stellar flux ratio as a continuous function of the spatial scale around emission peaks, which have been identified using a simple peak identification algorithm.\footnote{Note that the details of the peak selection do not significantly affect the results, see \citet{Kruijssen2018}.} After pruning the peak sample with a Monte-Carlo technique to avoid overlapping flux contributions from adjacent peaks, we measure the quantity
\begin{equation}
\label{eq:bias}
    \mathcal{B}(l_{\rm ap})=\frac{\mathcal{F}_{\rm H\alpha,tot}}{\mathcal{F}_{\rm CO,tot}}\frac{\sum_i \mathcal{F}_{{\rm CO},i}(l_{\rm ap})}{\sum_i \mathcal{F}_{{\rm H\alpha},i}(l_{\rm ap})} ,
\end{equation}
where $\mathcal{B}$ is the relative change of the CO-to-H$\alpha$ flux ratio, $\mathcal{F}_{\rm H\alpha,tot}$ and $\mathcal{F}_{\rm CO,tot}$ represent the total H$\alpha$ and CO fluxes in the maps, and $\mathcal{F}_{{\rm H\alpha},i}(l_{\rm ap})$ and $\mathcal{F}_{{\rm CO},i}(l_{\rm ap})$ represent the flux within an aperture of diameter $l_{\rm ap}$ around an emission peak $i$. In the summation, $i$ runs to the total number of emission peaks in the reference map $n_{\text{ref}}$ when evaluating equation~(\ref{eq:bias}) around young stellar emission peaks, and to the total number of emission peaks in the gas map $n_{\text{CO}}$ when evaluating equation~(\ref{eq:bias}) around gas emission peaks. The relative change of the CO-to-H$\alpha$ flux ratio as a function of the aperture size in equation~(\ref{eq:bias}) is a sensitive probe of the cloud-scale evolutionary time-scales, which can be inferred by fitting a simple model to the empirical measurement of equation~(\ref{eq:bias}). The framework is introduced in \citet{KL14} and the application of this methodology through the {\sc Heisenberg} code\footnote{The {\sc Heisenberg} code is publicly available at \url{https://github.com/mustang-project/Heisenberg}.} is explained in detail in \citet{Kruijssen2018}; we refer to these works for further details. {\sc Heisenberg} has been demonstrated to provide an accurate characterisation of the cloud lifecycle using a broad range of galaxy simulations \citep{Kruijssen2018, Fujimoto2019, Haydon2020b, Jeffreson2021, Keller2021, Semenov2021}. It has been used to measure molecular cloud lifetimes in more than sixty galaxies \citep{Kruijssen2019,Chevance2020,Zabel2020,Kim2021,Kim2022}, as well as the atomic cloud lifetime in the LMC (\citetalias{Ward2020_HI}) and the lifetime of embedded massive star-forming regions \citep[traced by 24$\mu$m,][]{Kim2021}.

Without further constraints, our statistical technique can be used to determine the {\it relative} time-scales of any two tracers that are linked through a common evolutionary timeline. In order to determine an {\it absolute} time-scale for either of the two tracers, the absolute visibility time-scale of the other tracer must be known. In this work, to determine the time-scales associated with molecular clouds in the LMC, we use H$\alpha$ emission as a reference, both with and without the presence of continuum emission. \citet{Haydon2020b,Haydon2020} applied {\sc Heisenberg} to simulated H$\alpha$ maps that were generated by combining a numerical simulation of a disc galaxy with stochastic stellar population synthesis models. This allowed them to measure the characteristic time-scales associated with the H$\alpha$ emission arising from young star-forming regions in galaxies, and to determine expressions for the H$\alpha$ emission time-scale as a function of the metallicity and the star formation rate surface density (reflecting the effect of stochastic sampling from the stellar initial mass function). Using equations 11 and 12 from \citet{Haydon2020} we determine time-scales for the reference images used in this work of $4.67^{+0.15}_{-0.34}$\,Myr (SHASSA continuum-subtracted H$\alpha$) and $8.5^{+1.0}_{-0.8}$\,Myr (MCELS H$\alpha$ including continuum emission).

The coverage of the MAGMA CO survey of the LMC is incomplete. The survey contains pointings towards 114 of the 272 molecular clouds reported in the NANTEN survey. However, the observing strategy of MAGMA prioritised the most massive molecular cloud complexes reported by the NANTEN survey, such that it accounts for $\sim 80$\% of the total CO flux \citep{TonyWong2011}. Given that {\sc Heisenberg} determines a flux-weighted average, this is the key number of interest and it is more important to account for total flux rather than the total number of emission peaks. However, this still presents a potential issue in areas where molecular clouds and H\,{\sc ii} regions are present but are not covered by the MAGMA survey. In order to avoid underestimating  the time-scale for which both H\,{\sc ii} regions and molecular clouds co-exist, we mask all regions towards which molecular clouds are detected with the NANTEN survey but do not fall within the coverage of the MAGMA survey. A circular mask of $\sim 250$\,pc in diameter (sufficient to fully cover the clouds) is chosen to mask these regions in all maps (CO and H$\alpha$), which are therefore excluded from the analysis. In Appendix~\ref{limitedCoverage}, we repeat the same measurements without masking these regions and with adding artificial molecular clouds in the omitted regions using the reported cloud fluxes, positions, and dimensions from the NANTEN survey. We find that the resulting time-scales are all consistent with the key results presented in Section~\ref{results}.

\begin{figure*}
	\includegraphics[width=0.49\linewidth]{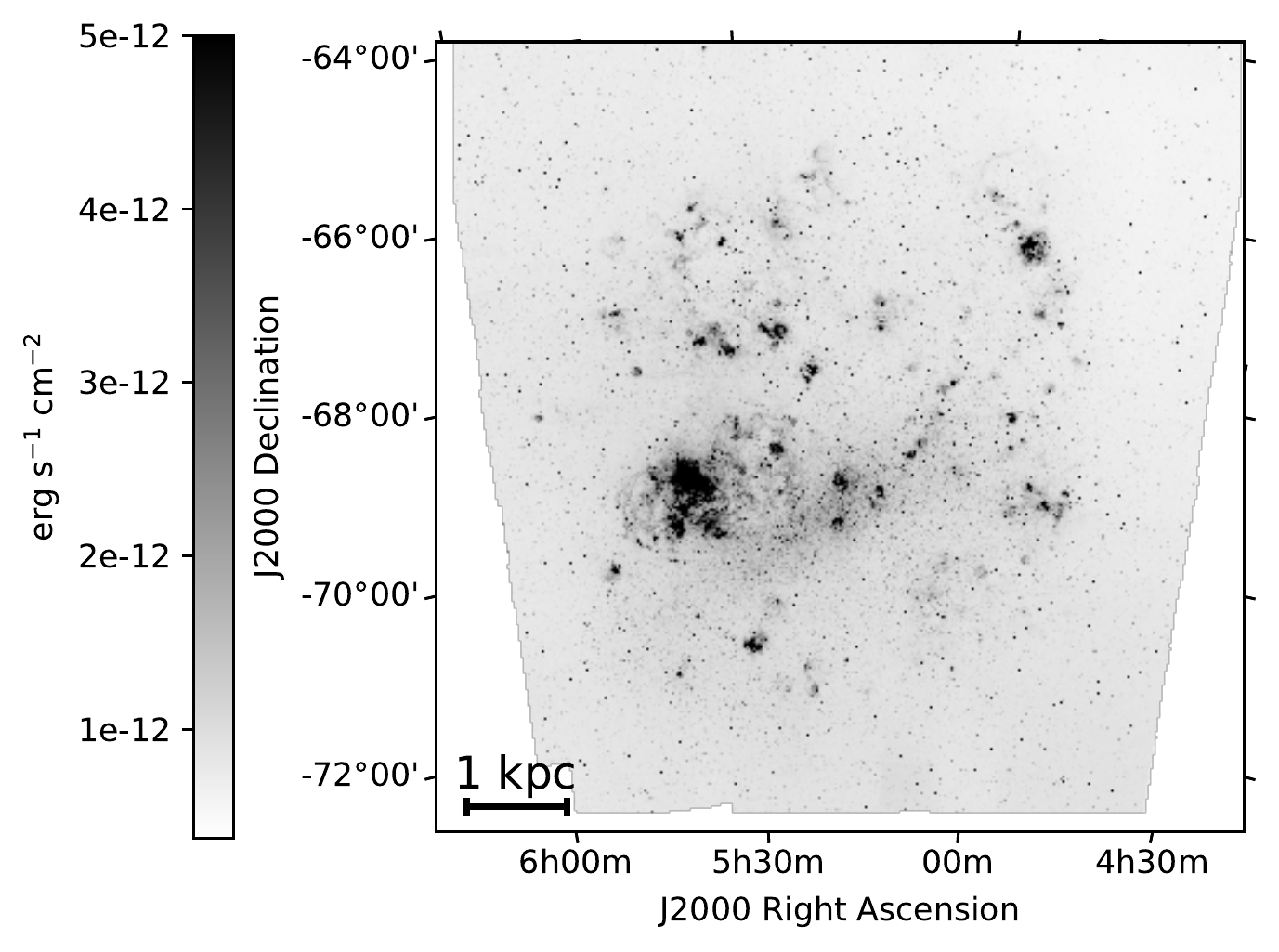}
	\includegraphics[width=0.49\linewidth]{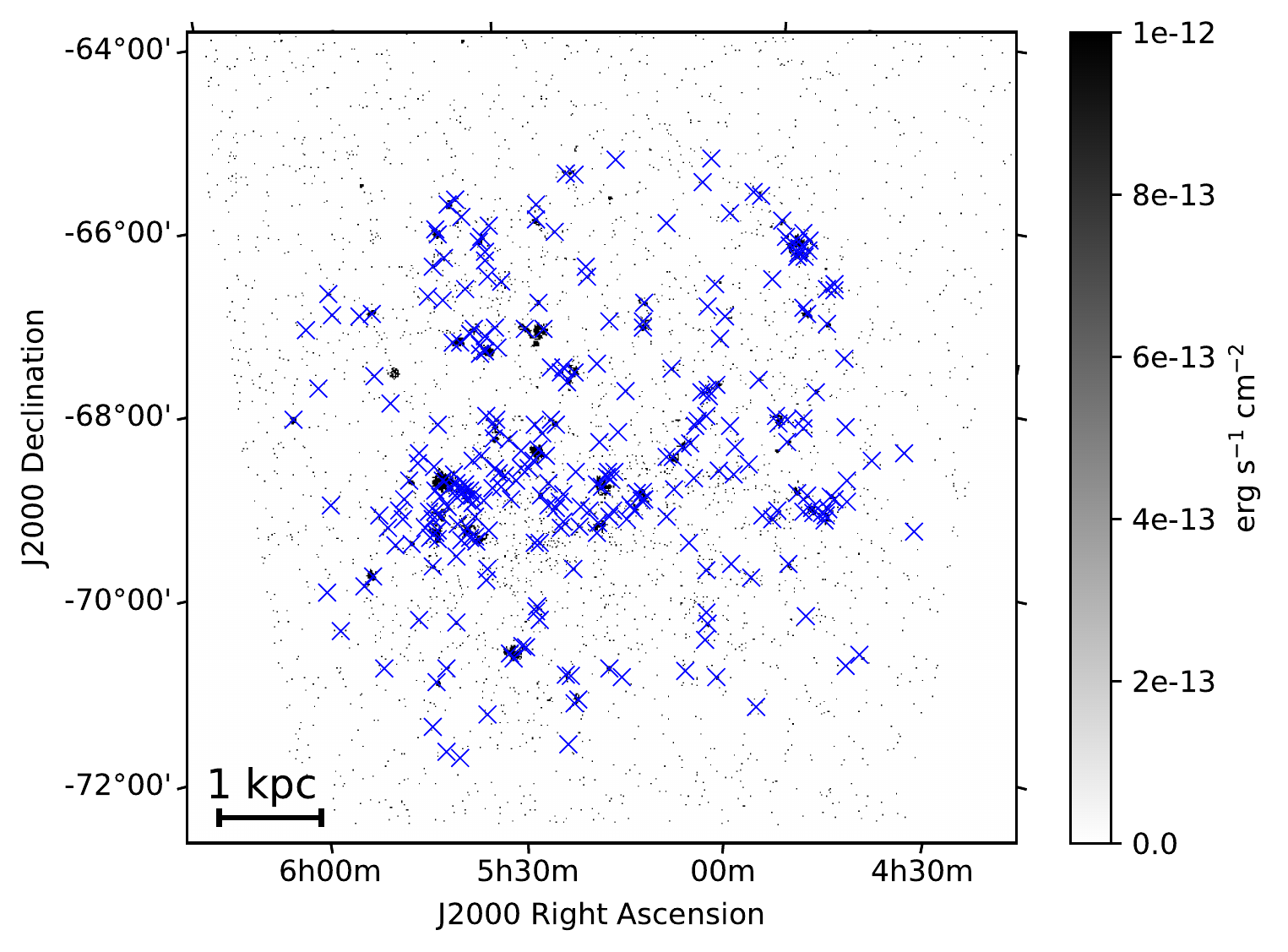}
	\includegraphics[width=0.49\linewidth]{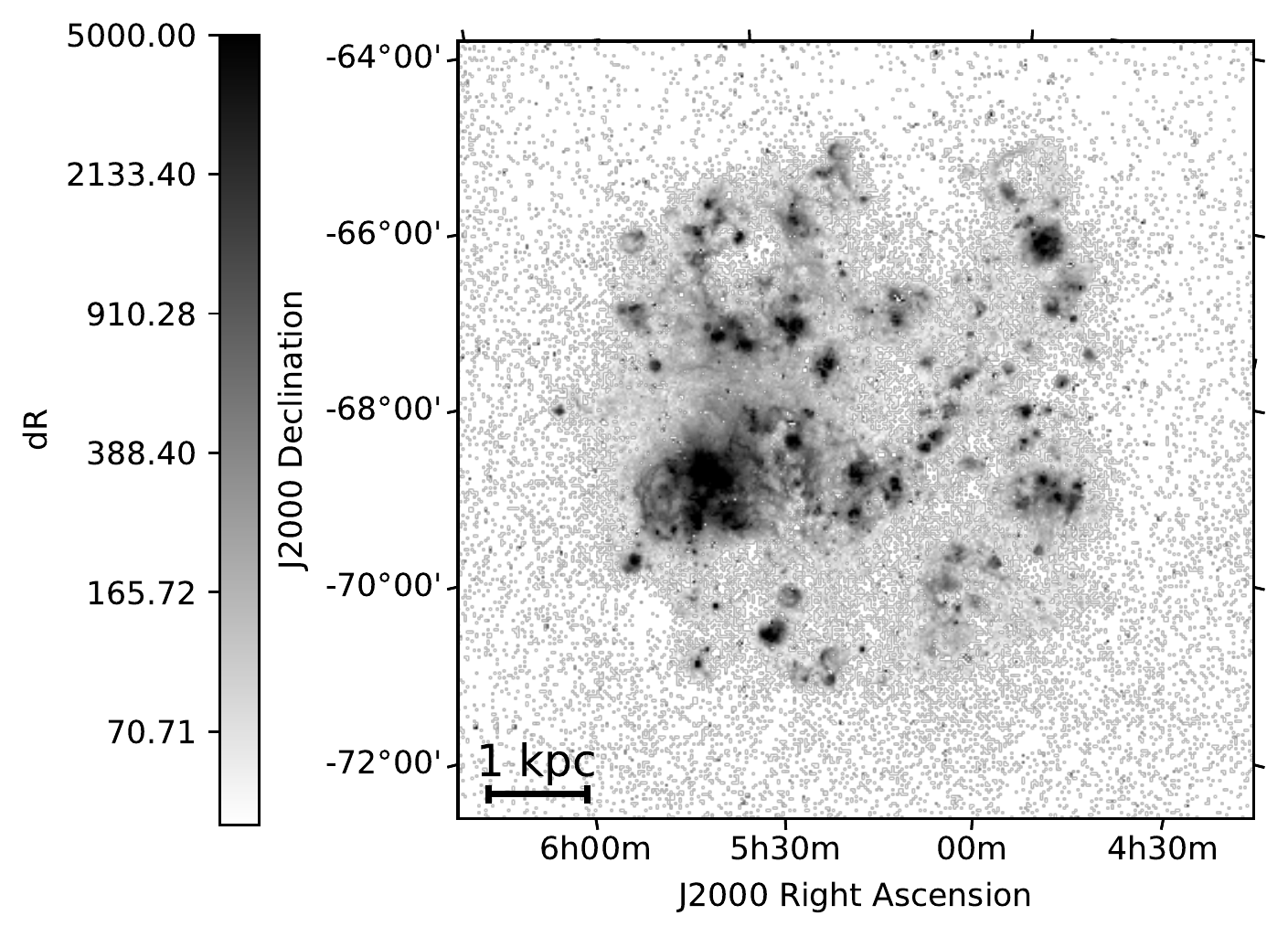}
	\includegraphics[width=0.49\linewidth]{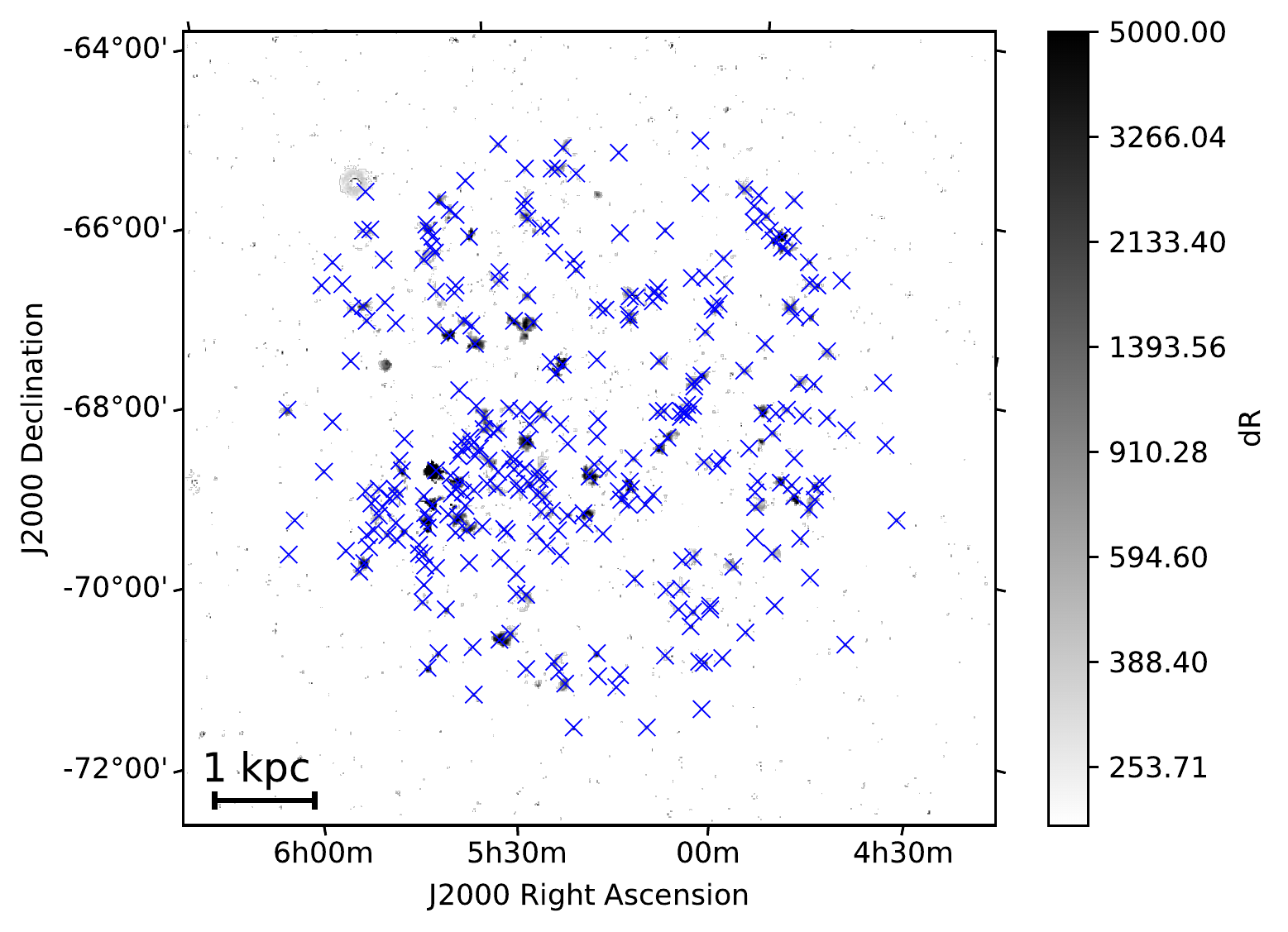}
	\includegraphics[width=0.49\linewidth]{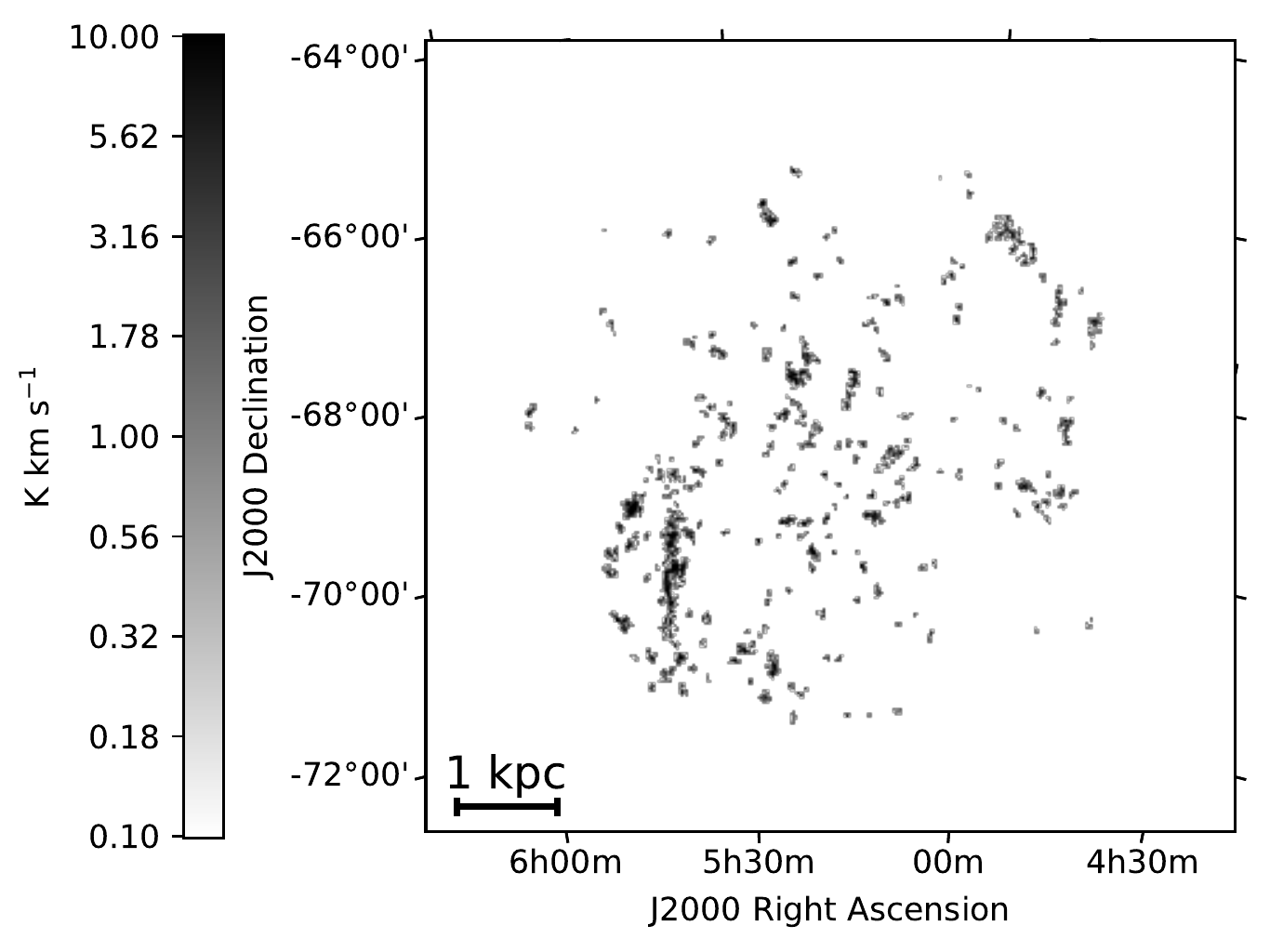}
	\includegraphics[width=0.49\linewidth]{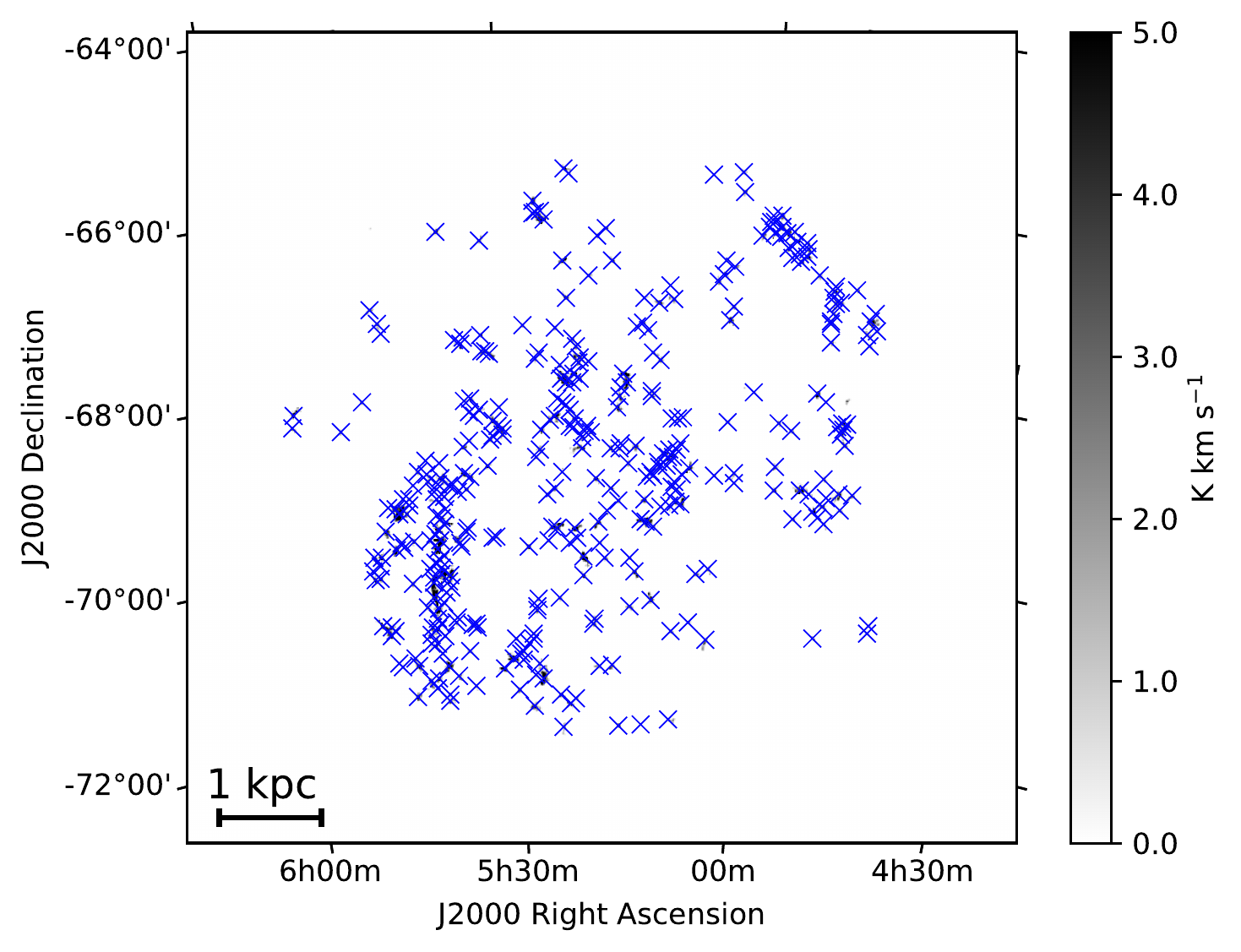}
	\caption{\label{fig:filteringImages} The images used in the main body of this work: MCELS H$\alpha$ including continuum (top), SHASSA continuum-subtracted H$\alpha$ (middle), and MAGMA CO (bottom), shown both before Fourier filtering (left) and after (right). In the bottom-left panel, the coverage of the observations closely traces the emission (see fig.~1 of \citealt{Wong2017}), implying that most of the pixels without emission are not covered by the MAGMA survey (see the text and Appendix~\ref{limitedCoverage} for details). In the right-hand panels, blue crosses indicate the identified emission peaks, around which the CO-to-H$\alpha$ flux ratio is measured as a function of spatial scale to constrain the evolutionary timeline of molecular cloud formation, evolution, and destruction (see the text for details).}
\end{figure*}

In order to remove diffuse emission that cannot be attributed to independent star-forming regions, we implement the iterative filtering in Fourier space outlined in \citet{Hygate2019}. We use a Gaussian shaped filter with a width of some length-scale, determined as a multiple of the characteristic separation length between independent regions $\lambda$ as obtained by the previous iteration of {\sc Heisenberg}. In order to achieve a good compromise between the removal of the diffuse emission and the preservation of the emission peaks, we select the lowest (integer) multiple of $\lambda$ ensuring that the corrective factor applied to compensate any over-subtraction caused by the shape of the Gaussian filter stays close to unity ($q_{\text{con}}>0.9$; see \citealt{HygatePhD}). In this case, we therefore apply a Gaussian filter of width $9\times \lambda$. For another recent summary of this approach, see appendix~A of \citet{Chevance2020}.

The H$\alpha$ and CO emission maps used in our analysis are shown in Figure~\ref{fig:filteringImages}, with the left column including the diffuse emission and the right column showing the maps after the diffuse emission has been filtered out. For reference, we also include the positions of the emission peaks around which the CO-to-H$\alpha$ flux ratio is evaluated as a function of the spatial scale.

\section{Results}

\label{results}

We now present our key measurements, such as the molecular cloud lifetime, as well as the time-scales associated with the destruction of molecular clouds in the LMC.

\subsection{De-correlation between CO and H$\alpha$ emission}
\label{decorrelation}
We apply the methodology described in Section~\ref{uncertaintyPrinciple} to two pairs of maps: the CO map and the SHASSA continuum-subtracted H$\alpha$ map, as well as the CO map and the MCELS H$\alpha$ map including continuum. Before proceeding to the basic measurement of equation~(\ref{eq:bias}), we point out that the flux-averaged nature of the adopted methodology can yield biased results when a single region in a galaxy is significantly brighter than the second-brightest region, because it violates the assumption of our methodology that the recent and imminent total star formation rate are approximately constant (see sect.~2.3 of \citealt{Kim2021} for a quantitative discussion). In the LMC, 30~Doradus qualifies as such a region, and in our fiducial analysis we therefore mask it using a circular aperture with a 200~pc radius, chosen to correspond to roughly twice the mean region separation length obtained with our measurements (see Section~\ref{lambda}). However, we do present the unmasked results alongside the fiducial case and show that the impact of 30~Doradus on our results is relatively minor (as was also shown in \citetalias{Ward2020_HI} for the atomic cloud lifetime).

\begin{figure*}
	\includegraphics[width=0.49\linewidth]{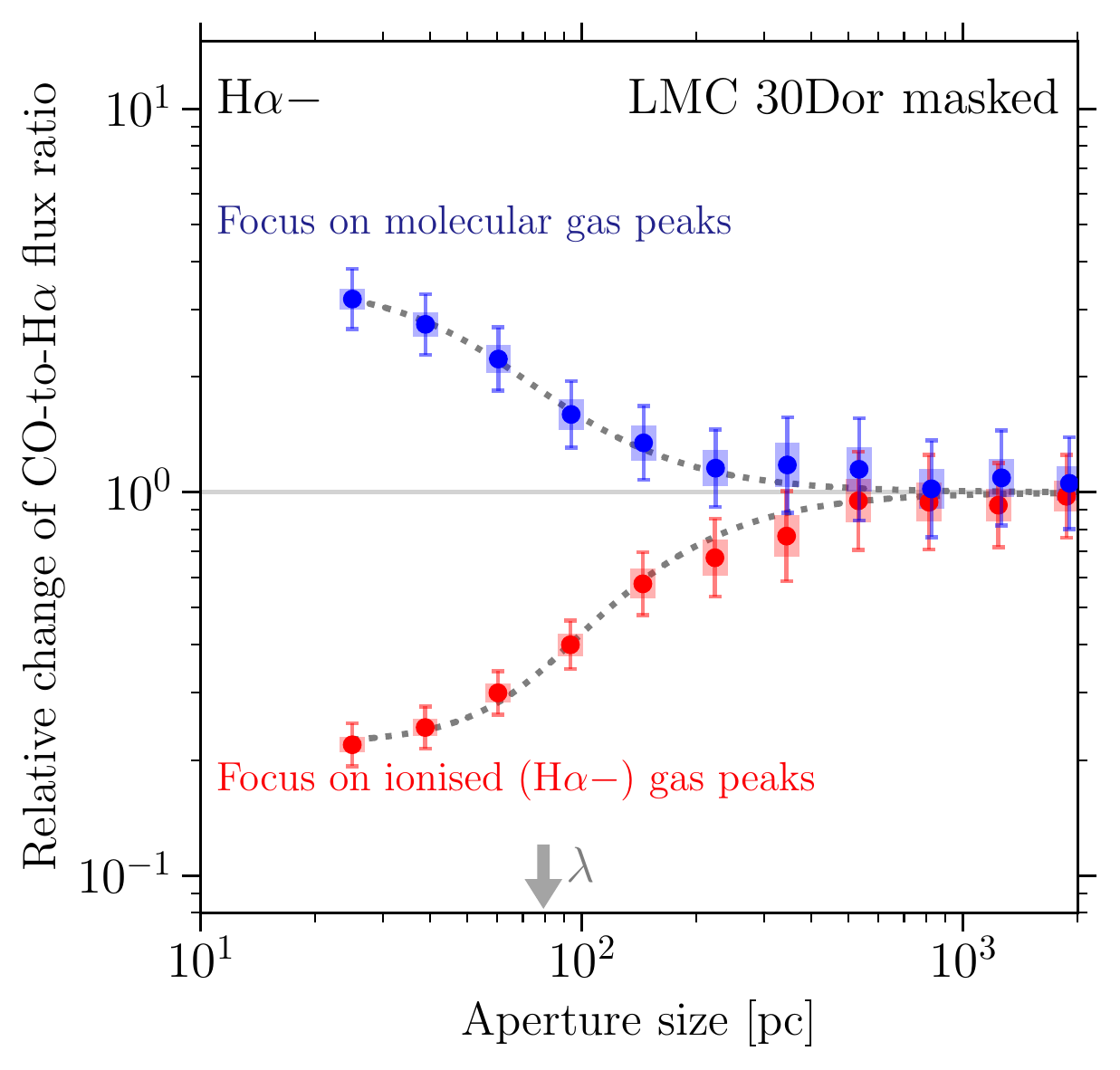}
	\includegraphics[width=0.49\linewidth]{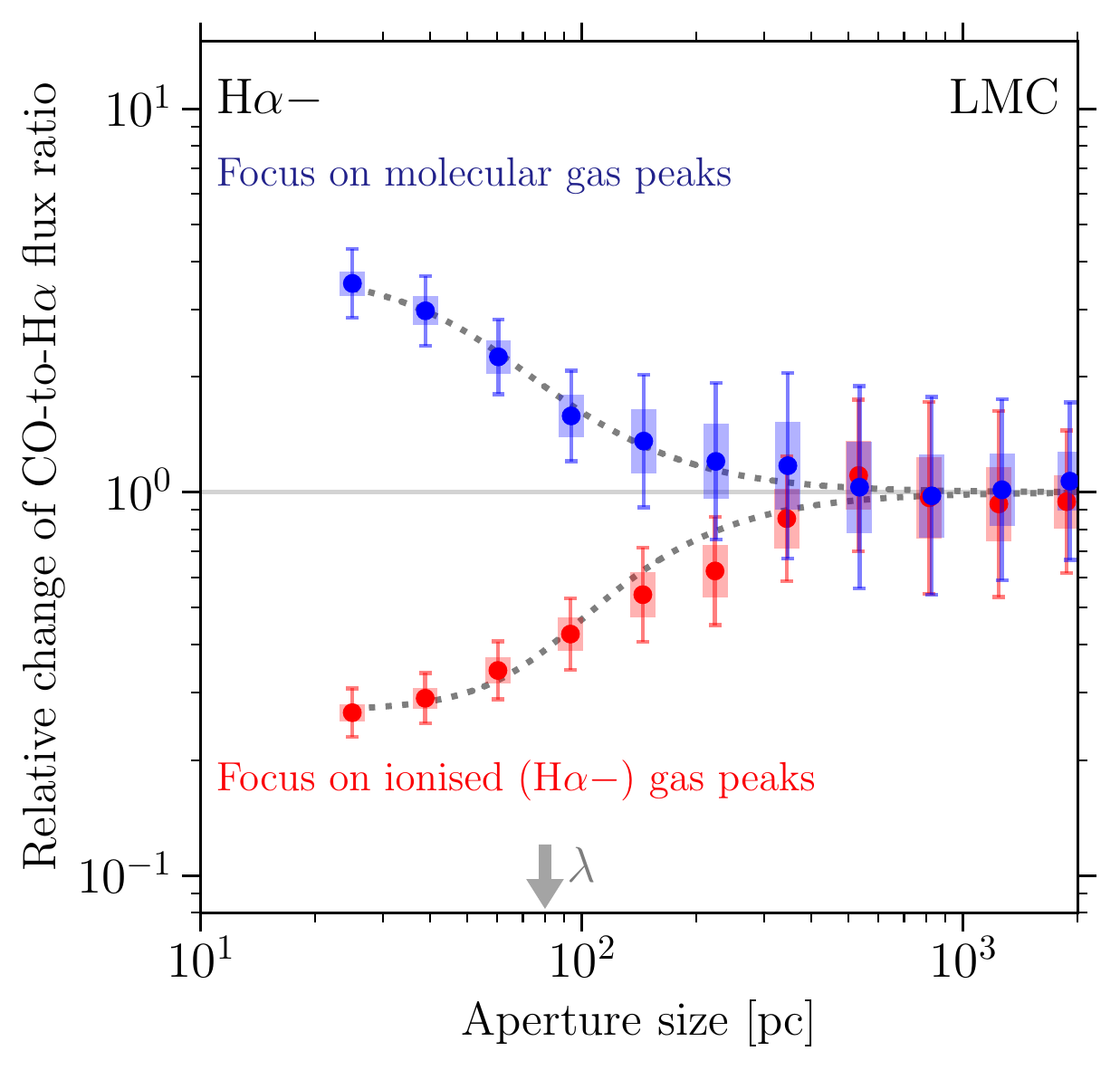}
	\includegraphics[width=0.49\linewidth]{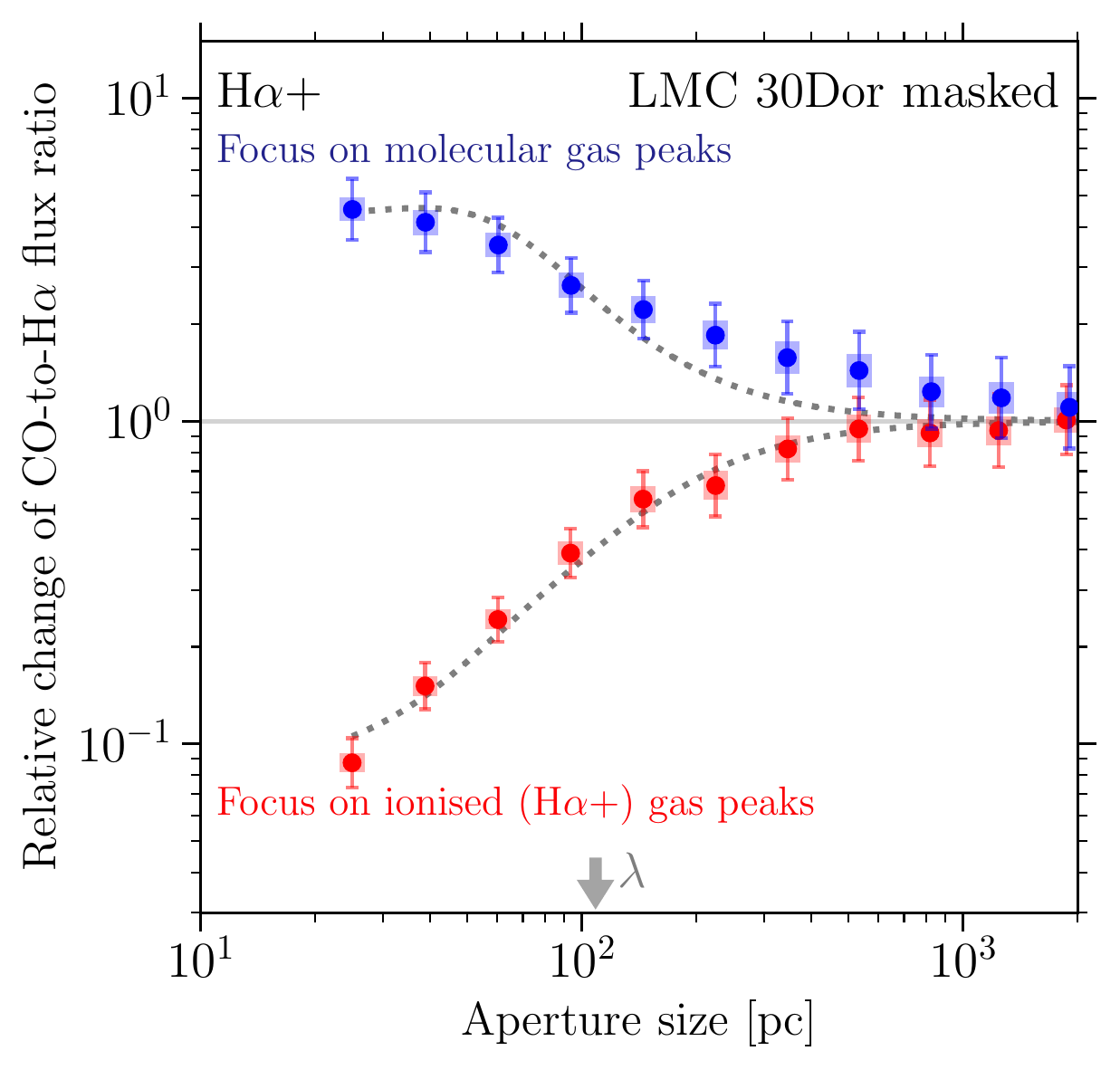}
	\includegraphics[width=0.49\linewidth]{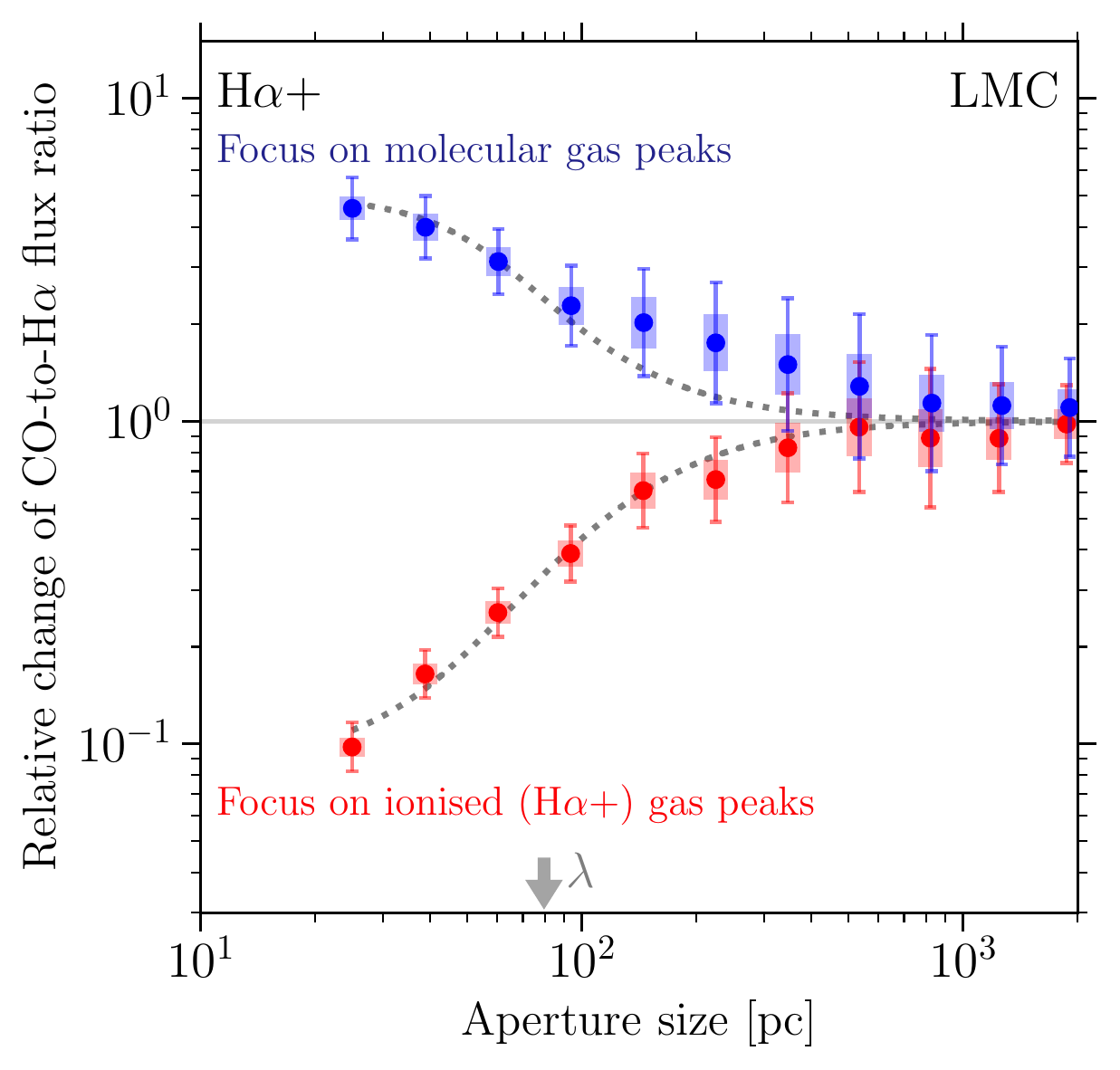}
	\caption{\label{fig:tuningForks} The relative change in the CO-to-H$\alpha$ emission flux ratio as a function of the size of the aperture within which that ratio is measured. Measurements focused on CO emission peaks are shown in blue and those focused on H$\alpha$ emission peaks are shown in red. Standard uncertainties are shown with the solid error bars, while uncertainties that take into account the covariance between the data are shown by the thick shaded bars. The model fitted to the data is indicated by the dotted curves and the measured mean separation length between independent regions ($\lambda$) is marked by the grey arrow. The results using the SHASSA continuum-subtracted H$\alpha$ image as the reference map are shown in the top row and those using the MCELS H$\alpha$ image that includes continuum emission are shown in the bottom row. The panels on the left show the fiducial case in which 30~Doradus is excluded from the analysis, while those on the right do include the region.}
\end{figure*}

Figure~\ref{fig:tuningForks} shows that the CO and H$\alpha$ emission are de-correlated on cloud scales ($\la100$~pc), i.e.\ focusing small apertures on peaks of CO emission leads to a deficit of H$\alpha$ emission compared to the galactic average, and vice versa. This is by now a well-known result that has been found across the local galaxy population \citep{Schruba2010,Kruijssen2019,Schinnerer2019,Chevance2020,Kim2021}, and it implies that (1) molecular clouds do not live much longer than the H{\sc ii} regions that they produce and (2) that the H{\sc ii} region appears late in the lifetime of a molecular cloud and destroys it rapidly \citep{Kruijssen2019,Chevance2021}. The fact that we find the same result here means that, qualitatively speaking, the same physical interpretation applies to molecular clouds in the LMC.

\subsection{Molecular cloud lifetimes in the LMC}

\label{fundamentalParameters}

We now turn to a quantitative discussion of the observables constrained by fitting the model of \citet{KL14} and \citet{Kruijssen2018} to the data points shown in Figure~\ref{fig:tuningForks}. A summary of our measurements is provided in Table~\ref{fun_tab}.

\begin{figure*}
	\includegraphics[width=0.49\linewidth]{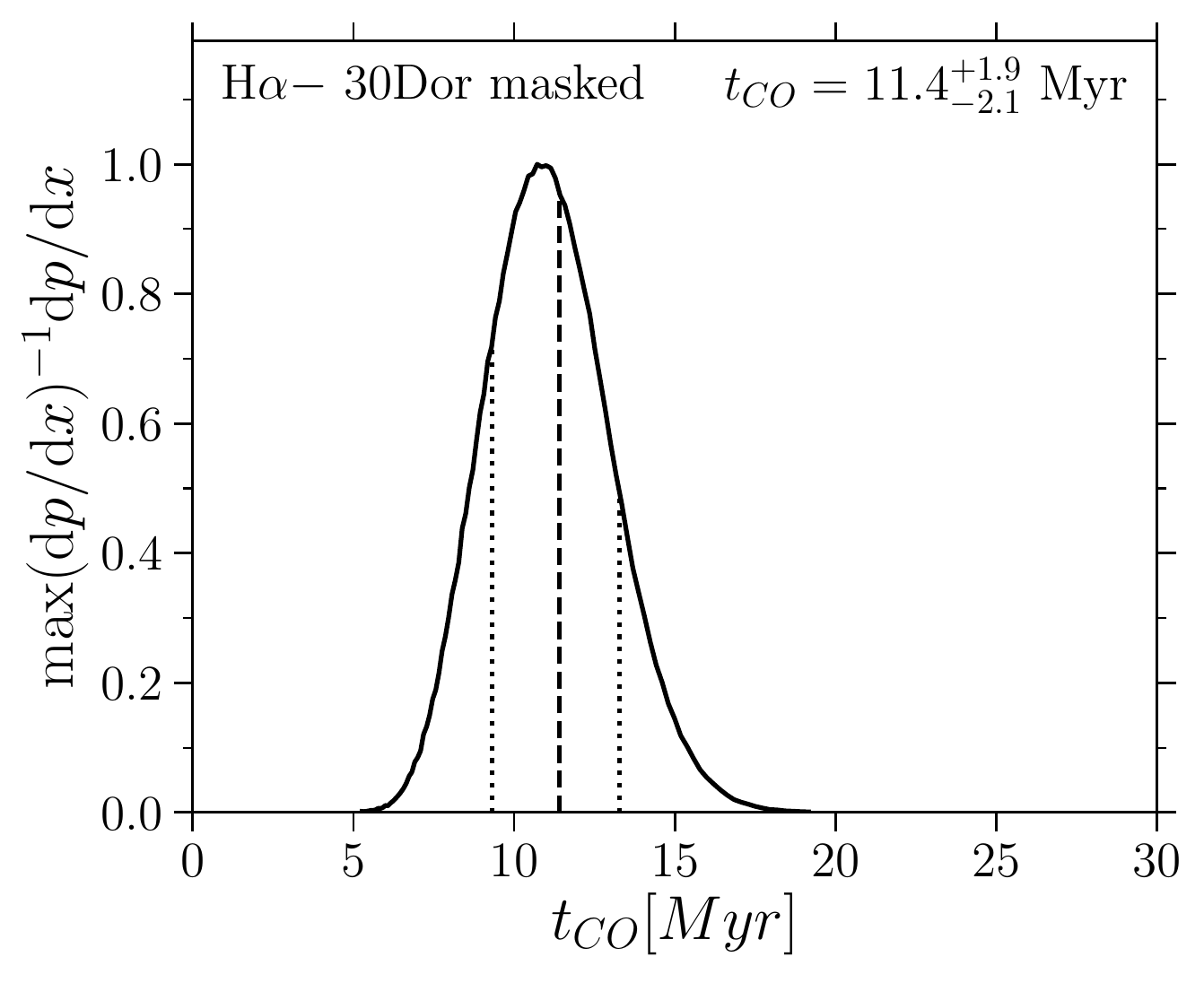}
	\includegraphics[width=0.49\linewidth]{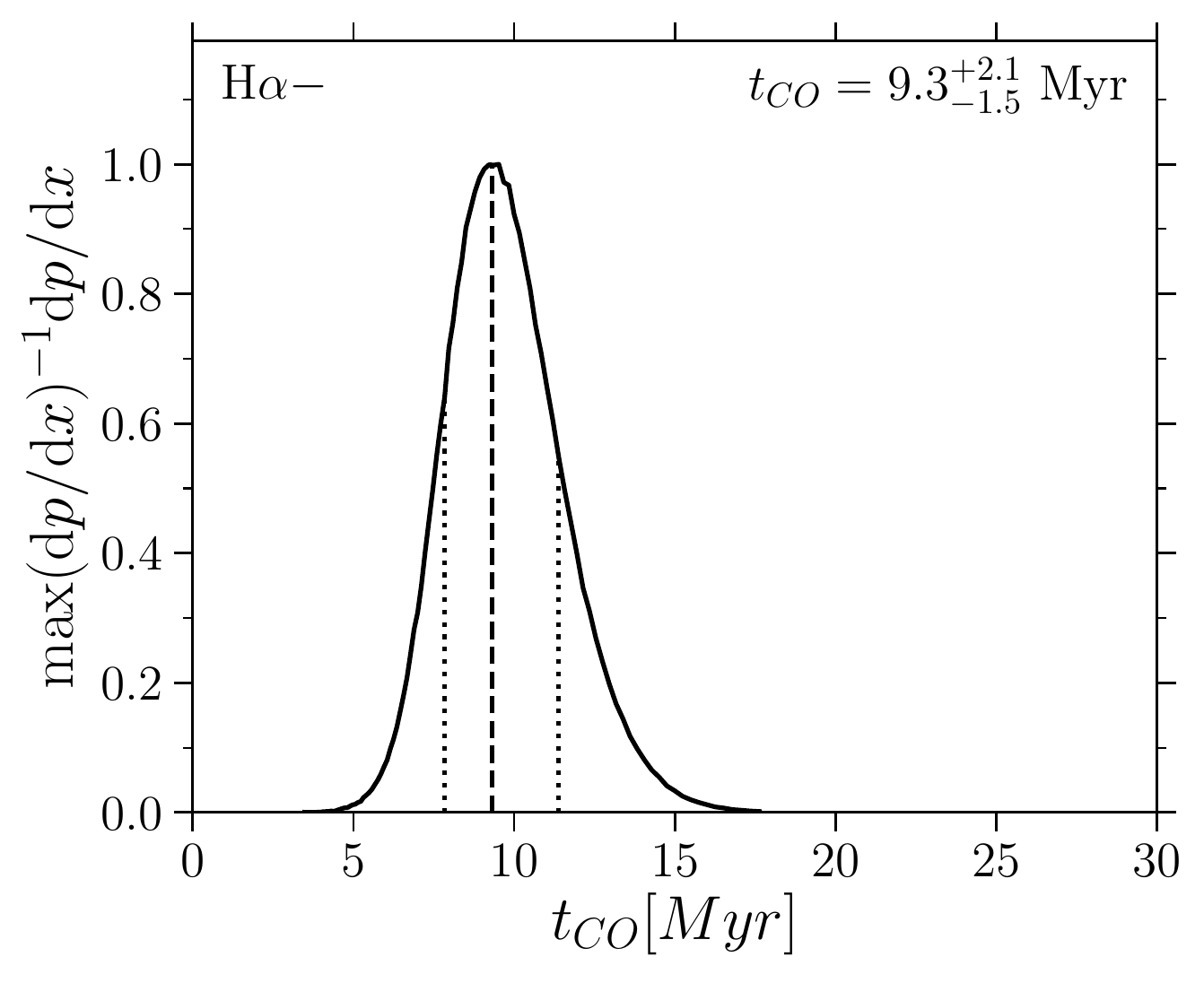}
	\includegraphics[width=0.49\linewidth]{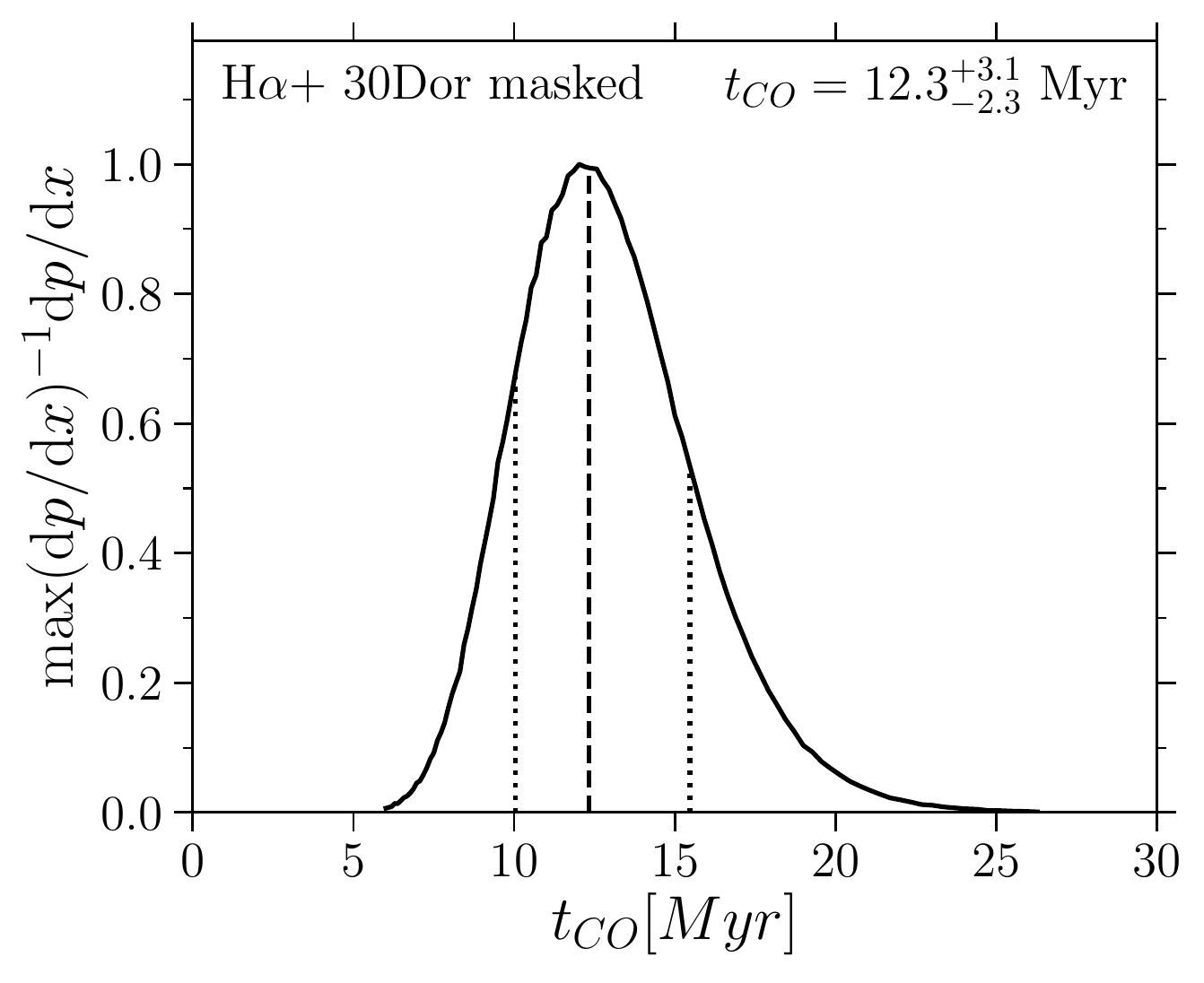}
	\includegraphics[width=0.49\linewidth]{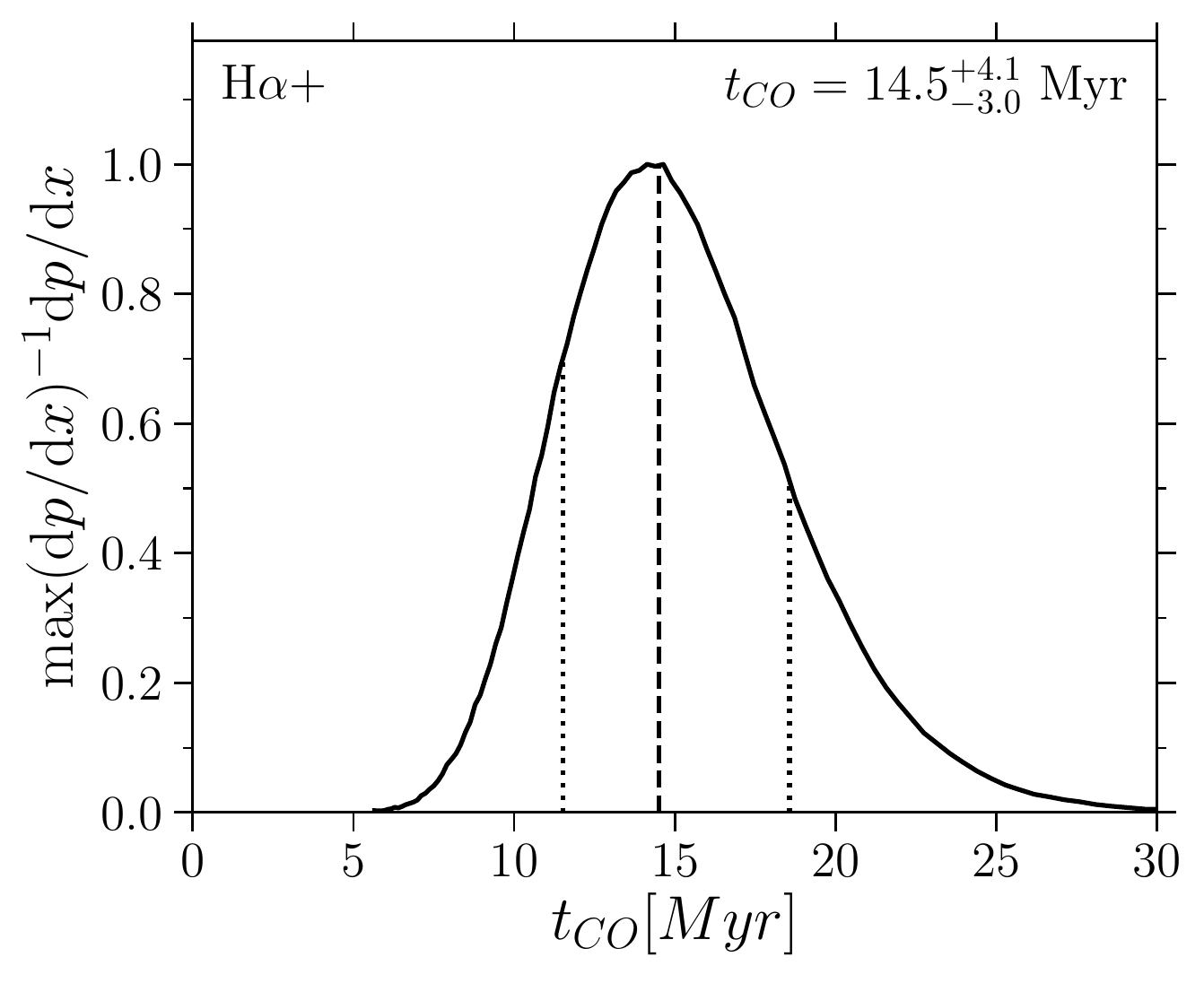}
	\caption{\label{fig:tgasPDFfig} PDFs of the measured molecular cloud lifetime ($t_{CO}$) in the LMC. Upper panels show PDFs derived using the SHASSA continuum-subtracted H$\alpha$ image as the reference map and lower panels show those using the MCELS H$\alpha$ image as the reference map. Panels on the left show the resulting PDFs with 30~Doradus excluded from the analysis, while those on the right do include 30~Doradus. Note that the measured values of the cloud lifetime using H$\alpha-$ and H$\alpha+$ converge when 30~Doradus is excluded from the analysis.}
\end{figure*}

The ratio between the molecular cloud lifetime and the reference time-scale provided by H$\alpha$ is encoded by the asymmetry between both branches in Figure~\ref{fig:tuningForks}. We can quantify the time-scale by fitting our model. When the SHASSA continuum-subtracted H$\alpha$ image (H$\alpha-$) is used as the reference map, a CO time-scale of 11.4$\substack{+1.9\\-2.1}$\,Myr is derived. Using the MCELS map containing both H$\alpha$ emission and continuum emission (H$\alpha+$), a time-scale of 12.3$\substack{+3.1\\-2.3}$\,Myr is measured. As discussed above, these fiducial measurements use maps in which the 30~Doradus region is omitted using a circular mask with a 200\,pc radius. When including 30~Doradus, we obtain time-scales of 9.3$\substack{+2.1\\-1.5}$\,Myr (SHASSA) and 14.5$\substack{+4.0\\-3.0}$\,Myr (MCELS).  The probability distribution functions (PDFs) of these four individual measurements of the molecular cloud lifetime are provided in Figure~\ref{fig:tgasPDFfig}. The time-scales calculated using different reference maps can be used to produce a combined time-scale. This is done using the procedure outlined in Appendix~\ref{combiningPDFs}, where we provide the resulting combined PDFs of the molecular cloud lifetime. The resulting combined time-scales are 11.8$\substack{+2.7\\-2.2}$\,Myr (fiducial, omitting 30~Doradus) and 12.8$\substack{+5.1\\-3.6}$\,Myr (including 30~Doradus). All measured molecular cloud lifetimes are listed in Table~\ref{fun_tab}.

Interestingly, when the region surrounding 30~Doradus is included, the time-scales derived using the different reference maps are inconsistent with one another. However, when 30~Doradus is omitted, the derived time-scales do converge. This shows that, while 30~Doradus has little impact on the measured atomic gas cloud lifetime (\citetalias{Ward2020_HI}), the effect on the measured molecular cloud time-scale is significant. In addition, we find that the impact of including 30~Doradus depends on whether continuum emission is included in the H$\alpha$ filter or not. These differences are likely due to the relative flux enclosed in the 30~Doradus region for different tracer maps (atomic or molecular gas, H$\alpha$ with or without continuum emission). In any case, this highlights very clearly the fact that the large flux from this single region biases our measurements, by violating the assumption of an approximately constant star formation rate (as discussed in Section~\ref{decorrelation} above and in sect.~2.3 of \citealt{Kim2021}). This further justifies masking 30~Doradus in our fiducial case. Therefore, we adopt a molecular cloud time-scale of $t_{\text{CO}}=11.8\substack{+2.7\\-2.2}$\,Myr, which is the time-scale obtained by combining the experiments in which 30~Doradus has been masked.

\begin{table*}
\caption{\label{fun_tab} Output parameters as derived by the H{\sc eisenberg} code using various images as reference maps. The results using the continuum-subtracted SHASSA H$\alpha$ image as a reference map are shown in the first and fourth column, while those using the MCELS H$\alpha$ image are shown in the second and fifth columns. The third and sixth columns show parameters derived by combining the results of the runs using the two H$\alpha$ images as reference maps. The results using the H{\sc i} column density map of the LMC \citep{Kim2003} as a reference map with the associated time-scale derived in \citetalias{Ward2020_HI} as the reference time-scale is given in the final column. For the experiments using an H$\alpha$ reference map, the measurements obtained when masking 30~Doradus with a 200\,pc mask are shown in the first three columns, whereas 30~Doradus is included in the next three columns. For consistency with our fiducial measurements, the same 200\,pc mask is also adopted for the measurements that use H{\sc i} emission as a reference in the final column.}
\begin{tabular}{l c c c c c c c}
\hline
 & H$\alpha$- & H$\alpha$+ & combined & H$\alpha$- & H$\alpha$+ & combined & H{\sc i} \\
$t_{\text{ref}}$ [Myr] & 4.67$\substack{+0.15\\-0.34}$ & 8.54$\substack{+0.97\\-0.82}$ & - & 4.67$\substack{+0.15\\-0.34}$ & 8.54$\substack{+0.97\\-0.82}$ & - & 48$\substack{+13\\-8}$ \\
30~Doradus mask & 200\,pc & 200\,pc & 200\,pc & No mask & No mask & No mask & 200\,pc \\ \hline
$n_{\text{ref}}$ & 297 & 276 & - & 299 & 285 & - & 827 \\
$n_{\text{CO}}$ & 347 & 342 & - & 358 & 352 & - & 353 \\
$\chi^{2}$ & 0.28 & 2.10 & - & 0.25 & 0.85 & - & 0.79 \\
$t_{\text{CO}}$ [Myr] & 11.4$\substack{+1.9\\-2.1}$ & 12.3$\substack{+3.1\\-2.3}$ & 11.8$\substack{+2.7\\-2.2}$ & 9.3$\substack{+2.1\\-1.5}$ & 14.5$\substack{+4.0\\-3.0}$ & 12.8$\substack{+5.1\\-3.6}$ & 8.4$\substack{+2.1\\-2.0}$ \\
$t_{\text{fb}}$ [Myr] & 1.1$\substack{+0.2\\-0.2}$ & 1.4$\substack{+0.5\\-0.3}$ & 1.2$\substack{+0.3\\-0.2}$ & 1.1$\substack{+0.3\\-0.3}$ & 1.0$\substack{+0.7\\-0.5}$ & 1.1$\substack{+0.4\\-0.4}$ & 2.8$\substack{+1.0\\-1.2}$ \\
$\lambda$ [pc] & 79$\substack{+18\\-13}$ & 109$\substack{+22\\-13}$ & 92$\substack{+20\\-13}$ & 80$\substack{+23\\-15}$ & 80$\substack{+23\\-11}$ & 80$\substack{+23\\-14}$ & 68$\substack{+8\\-7}$ \\
$r_{\text{ref}}$ [pc] & 15.3$\substack{+0.5\\-0.7}$ & 9.7$\substack{+0.2\\-0.1}$ & - & 16.0$\substack{+1.2\\-1.1}$ & 10.2$\substack{+0.3\\-0.3}$ & - & - \\
$r_{\text{CO}}$ [pc] & 13.2$\substack{+0.6\\-0.4}$ & 14.7$\substack{+0.7\\-0.7}$ & - & 13.4$\substack{+0.7\\-0.5}$ & 13.6$\substack{+0.8\\-0.4}$ & - & - \\
$v_{\text{fb}}$ [km s$^{-1}$] & 12$\substack{+2\\-2}$ & 11$\substack{+3\\-2}$ & - & 12$\substack{+4\\-2}$ & 13$\substack{+14\\-5}$ & - & - \\
$\epsilon_{\text{SF}}$ [\%] & 3.7$\substack{+1.2\\-1.0}$ & 3.7$\substack{+1.5\\-1.0}$ & - & 3.0$\substack{+1.1\\-0.8}$ & 4.7$\substack{+2.0\\-1.4}$ & - & - \\
     \hline
\end{tabular}
\end{table*}


\subsection{The destruction of molecular clouds}

\label{feedbackVelocity}

\begin{figure*}
	\includegraphics[width=0.49\linewidth]{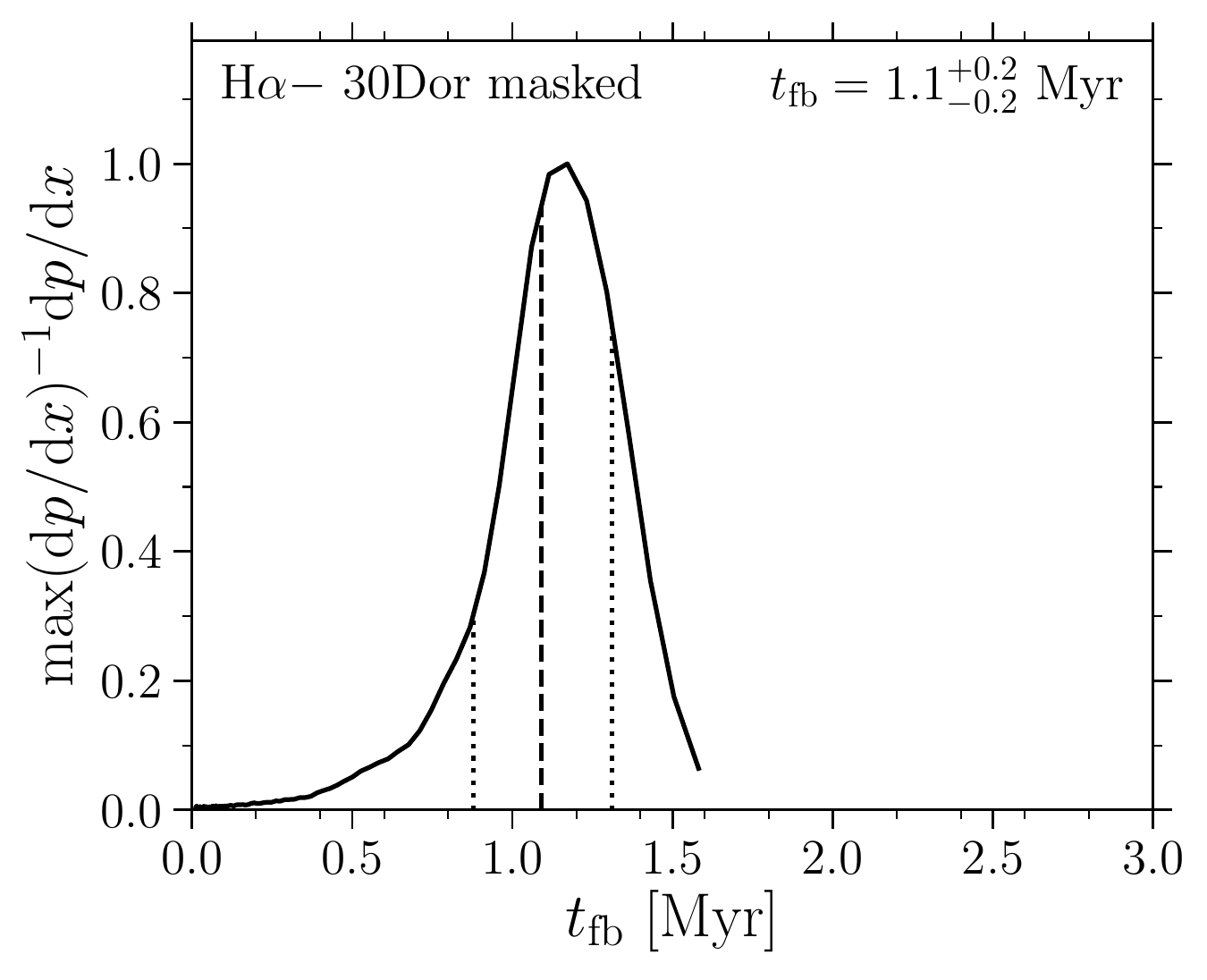}
	\includegraphics[width=0.49\linewidth]{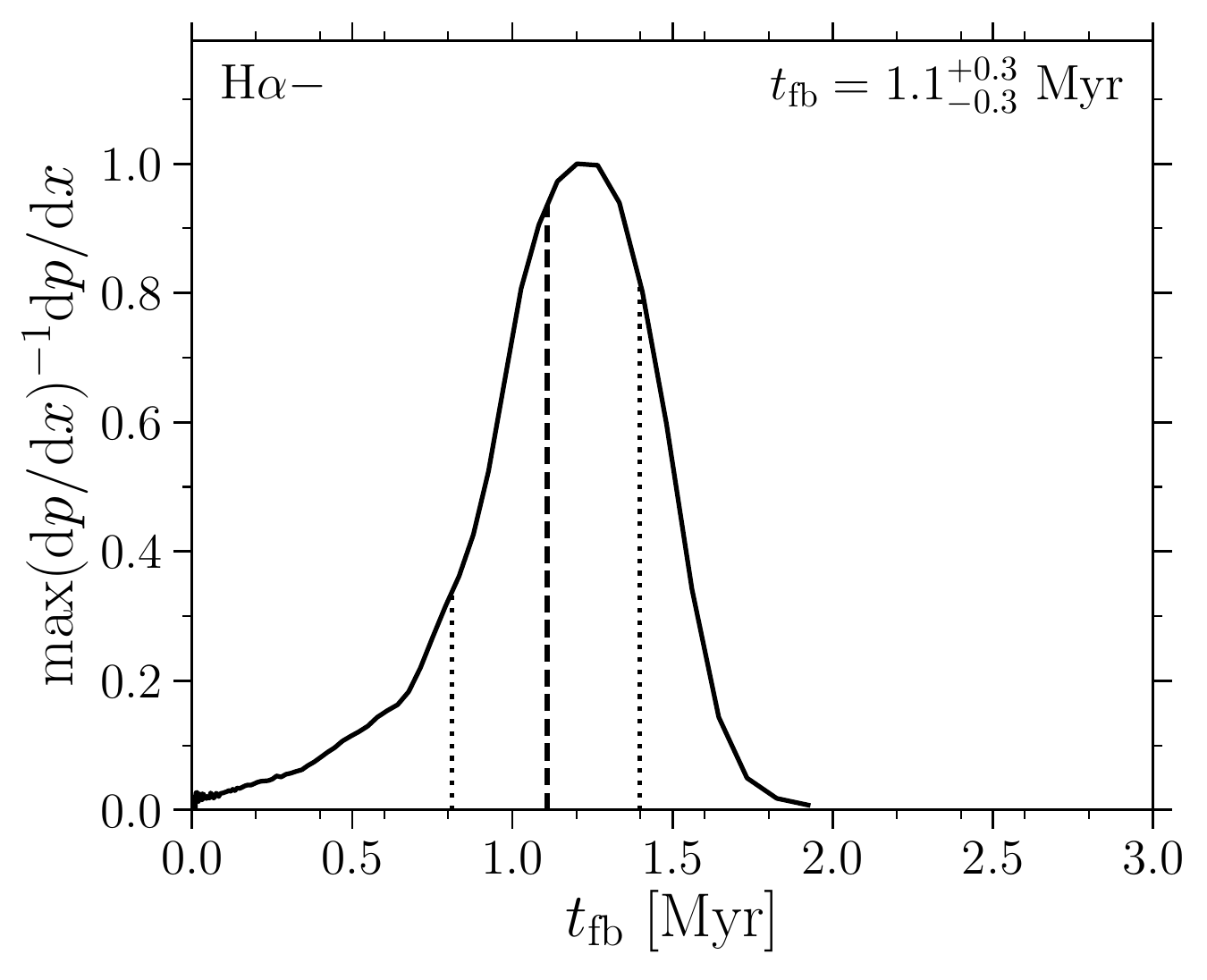}
	\includegraphics[width=0.49\linewidth]{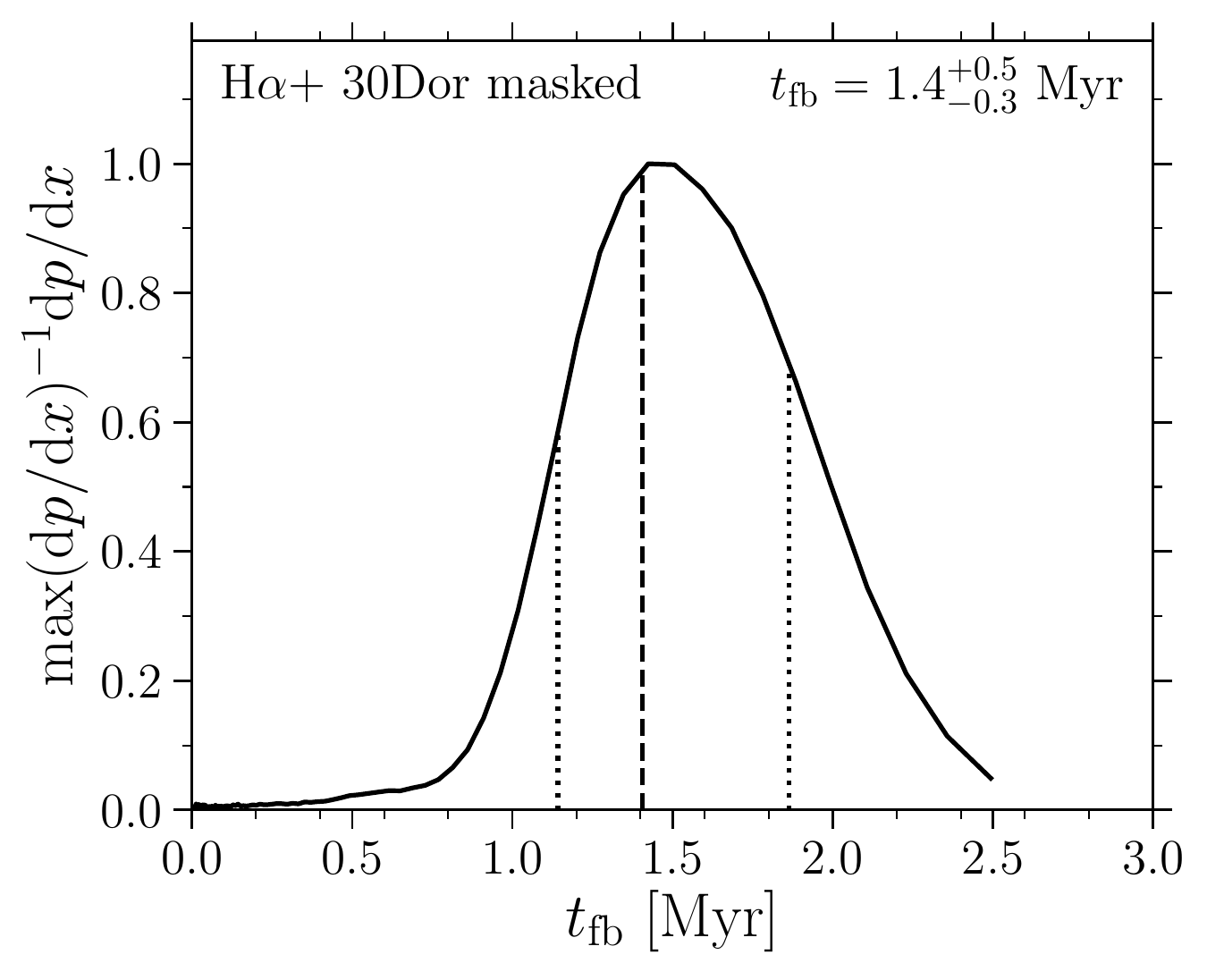}
	\includegraphics[width=0.49\linewidth]{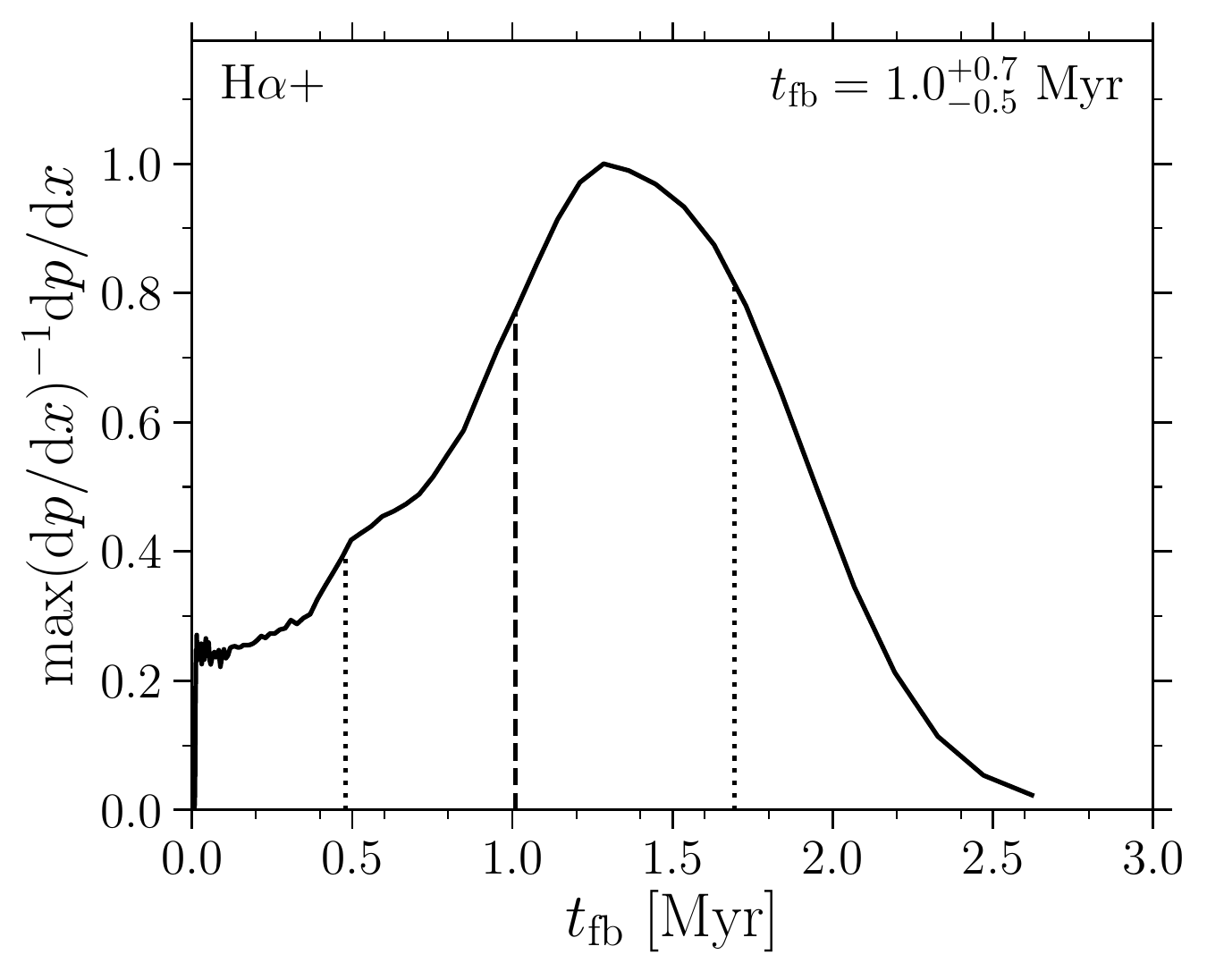}
	\caption{\label{fig:toverPDFfig} PDFs of the measured overlap time-scale for which H$\alpha$ and CO emission co-exist ($t_{\text{fb}}$) in the LMC. Upper panels show PDFs derived using the SHASSA continuum-subtracted H$\alpha$ image as the reference map and lower panels show those using the MCELS H$\alpha$ image as the reference map. Panels on the left show the resulting PDFs with 30~Doradus excluded from the analysis, while those on the right do include 30~Doradus.}
\end{figure*}

In addition to the average lifetime of CO emitting clouds in the LMC, {\sc Heisenberg} also determines the time-scale for which CO emission and H$\alpha$ emission co-exist. This time-scale is encoded by the degree of flattening of both branches at small aperture sizes in Figure~\ref{fig:tuningForks}. Assuming that molecular clouds are dispersed by feedback from massive stars and that the process begins with the emergence of H\,{\sc ii} regions, this overlap time-scale corresponds to the feedback time-scale, i.e.~the time taken for feedback to destroy the molecular clouds. For the purposes of this section, we assume that the overlap time-scale between H$\alpha$ emission and CO emission is representative of the average dispersal time of a molecular cloud in the LMC (this line of reasoning is discussed in more detail in \citealt{Chevance2020}). We measure feedback time-scales of 1.0--1.4\,Myr and 1.1\,Myr for the H$\alpha$+ and H$\alpha$- runs, respectively, depending on whether the immediate surroundings of 30~Doradus are masked (see Table~\ref{fun_tab} and Figure~\ref{fig:toverPDFfig}). While there is a small spread between these time-scales, it is well within the stated uncertainties in Table~\ref{fun_tab}. These consistently short feedback time-scales suggest that the dispersal of molecular clouds in the LMC is dominated by early feedback processes that precede the first supernovae, which are typically expected at ages of at least 4\,Myr \citep[e.g.][]{Leitherer2014}, considerably longer than the feedback time-scales inferred here.

Also measured by {\sc Heisenberg} are the average radii of the H$\alpha$ emission and CO emission peaks ($r_{\rm ref}$ and $r_{\rm CO}$ respectively), presented in Table~\ref{fun_tab}. We can use the derived feedback time-scale and the CO cloud radius to calculate the velocity at which gas is cleared from the region surrounding the newly formed star as $v_{\text{fb}} = r_{\text{CO}} / t_{\text{fb}}$. From each of the separate {\sc Heisenberg} runs presented in this paper, we derive an average feedback velocity of 12$\pm$2\,km\,s$^{-1}$. This is consistent with direct observations of the expansion velocities of H\,{\sc ii} regions in the LMC \citep[e.g.][]{Ambrocio-Cruz2016,Ward2016,McLeod2019}, adding further credence to recent findings that pre-supernova feedback dominates molecular cloud destruction \citep{Chevance2021}. By contrast, the expansion velocities associated with supernova remnants tend to be significantly higher, in the range 25--50\,km\,s$^{-1}$ \citep{Ambrocio-Cruz2016}. 

\subsection{Spatial distribution of molecular clouds and star-forming regions in the LMC}

\label{lambda}

\begin{figure*}
	\includegraphics[width=0.49\linewidth]{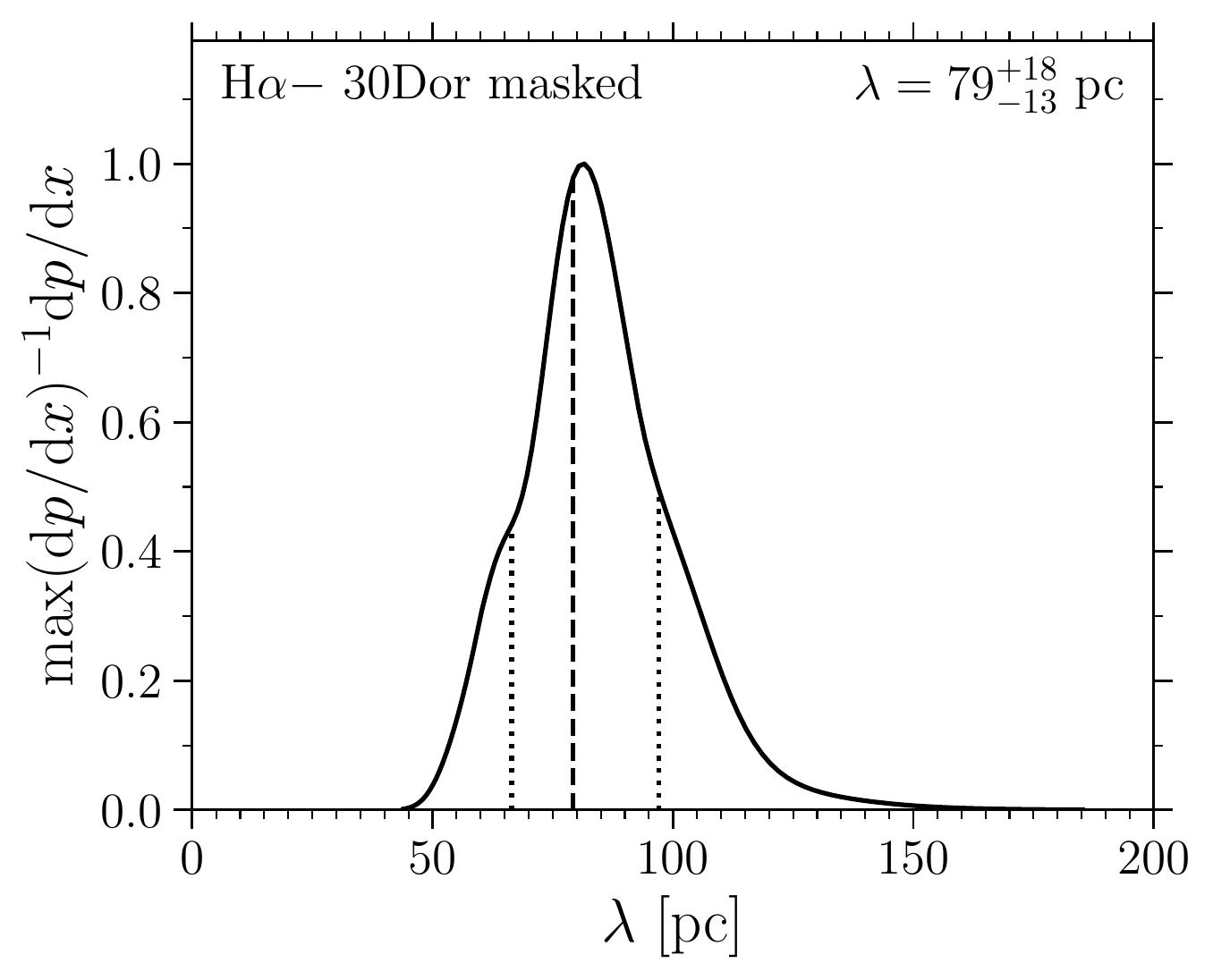}
	\includegraphics[width=0.49\linewidth]{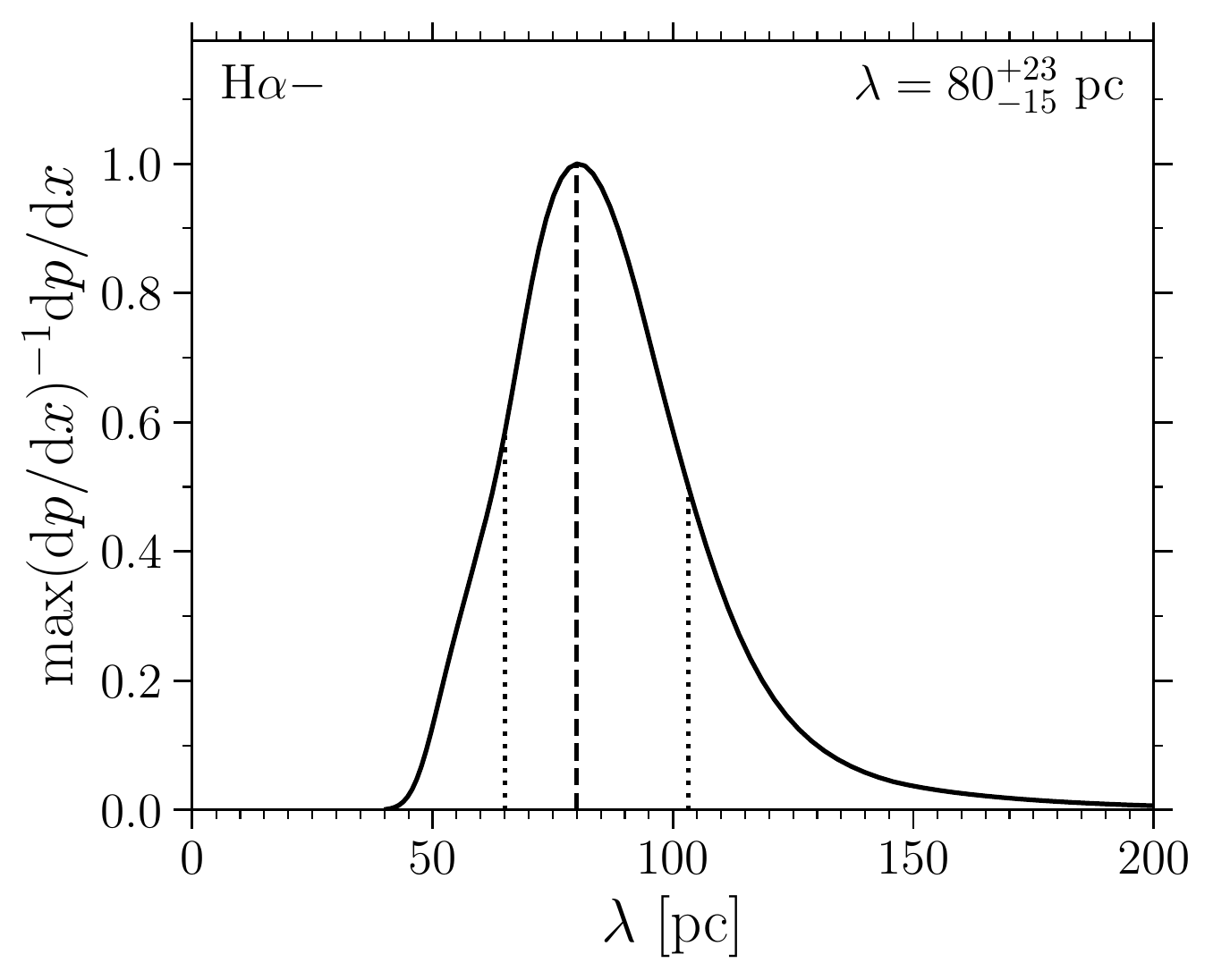}
	\includegraphics[width=0.49\linewidth]{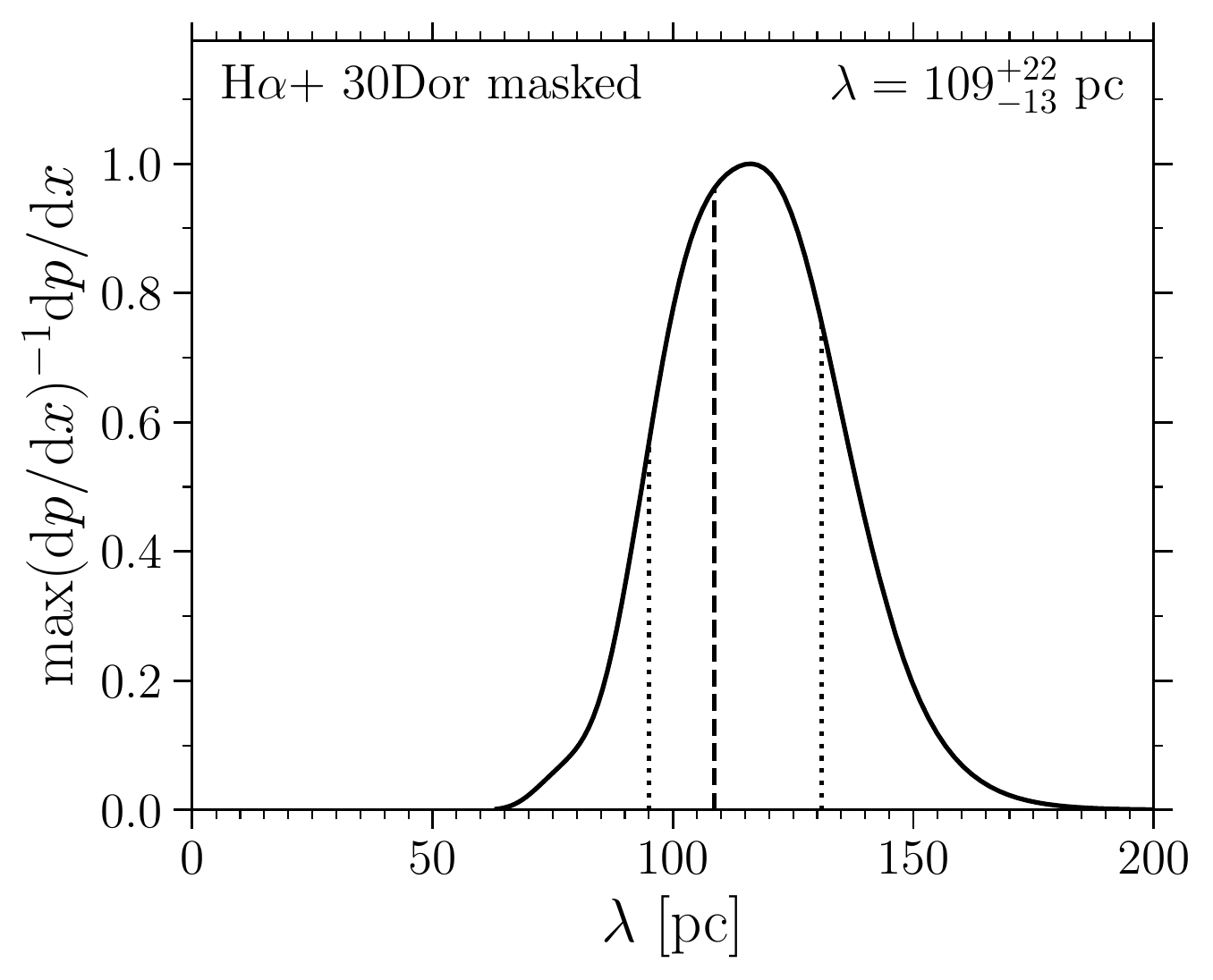}
	\includegraphics[width=0.49\linewidth]{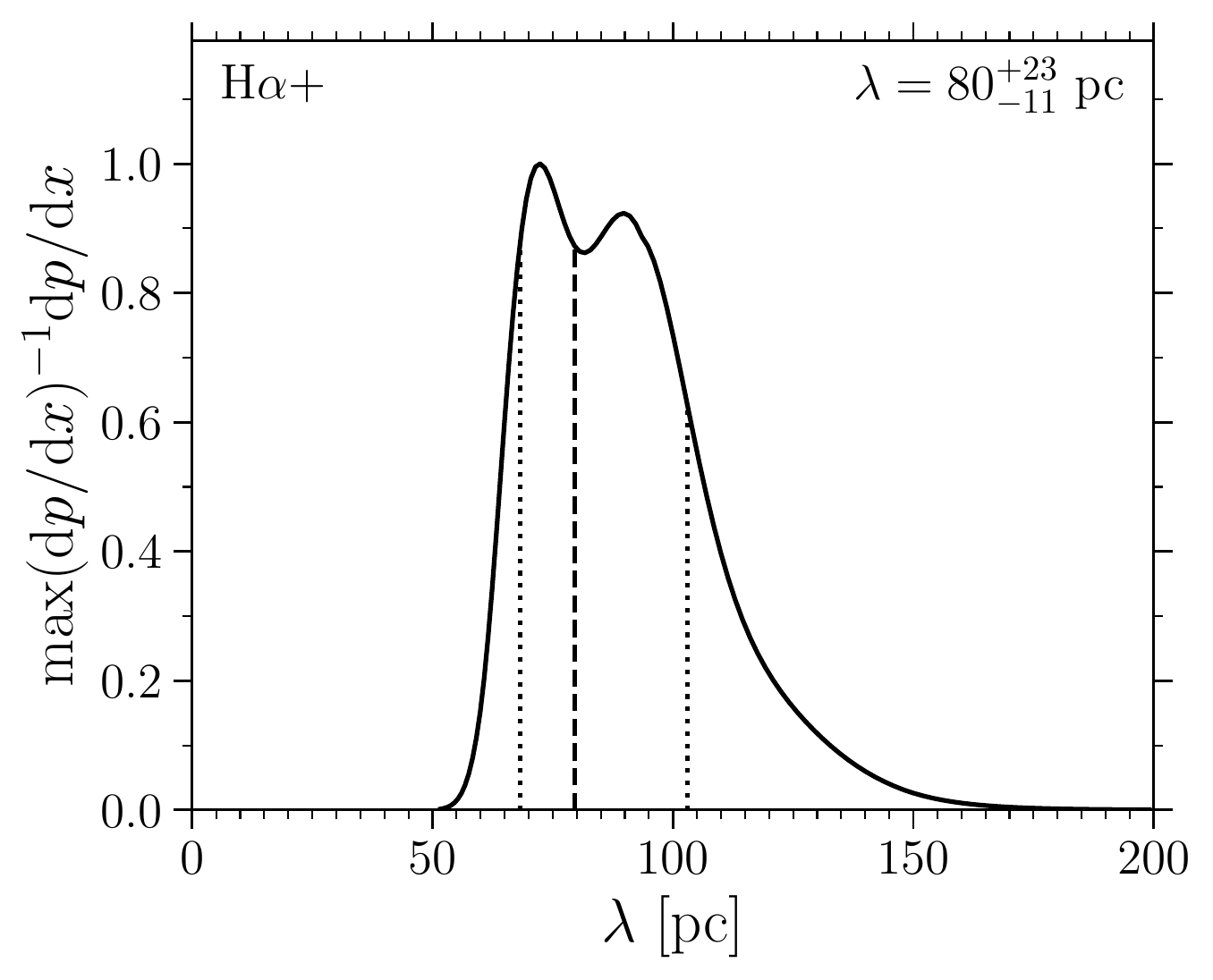}
	\caption{\label{fig:lambdaPDFfig} PDFs of the measured characteristic separation length ($\lambda$) in the LMC. Upper panels show PDFs derived using the SHASSA continuum-subtracted H$\alpha$ image as the reference map and lower panels show those using the MCELS H$\alpha$ image as the reference map. Panels on the left show the resulting PDFs with 30~Doradus excluded from the analysis, while those on the right do include 30~Doradus.}
\end{figure*}

The characteristic separation length between independent regions of star formation, $\lambda$, is shown in Table~\ref{fun_tab} and Figure~\ref{fig:lambdaPDFfig} for the set of experiments performed in this work. Values of $\lambda \sim 80$\,pc are measured in three of the four measurements in which we use H$\alpha$ image as the reference map; only the H$\alpha-$ run with 30~Doradus masked results in a different separation length of 109$\substack{+22\\-13}$\,pc. Despite this slight difference, we note that the results from all four runs are consistent within the error bars. As shown by Figure~\ref{fig:tuningForks}, these separation lengths are much larger than the working resolution of 25\,pc, implying that they are well-resolved.

\citet{Chevance2020} found that the length scale on which H\,{\sc ii} regions and molecular clouds de-correlate and undergo independent lifecycles, $\lambda$, spans a range of $\sim 100-250$\,pc across a sample of nine nearby galaxies. To within the uncertainties, the LMC is fully consistent with this range of characteristic separation lengths. Based on a similar analysis of the nearby flocculent galaxy NGC300, \citet{Kruijssen2019} suggest that the characteristic separation between independent regions closely matches the gas disc scale height, supporting the idea that the ISM is structured by feedback from young stellar regions. In the LMC, the atomic gas disc scale height has been estimated to 180\,pc by \cite{Padoan2001}. While the molecular gas disc scale height in the LMC has not been measured, \citet{Dawson2013} adopt a fiducial value of 90~pc, arguing that the molecular gas scale height is approximately a factor of two smaller than the atomic gas scale height. If this rough estimate is correct, the LMC is the second galaxy after NGC300 where the region separation length matches the molecular gas disc scale height.

\section{Discussion}

\label{discussion}

In this section, we place our results in the context of our previously measured atomic cloud lifetimes in the LMC and use the measured atomic and molecular cloud lifetimes to constrain the physical mechanisms driving the cloud lifecycle. Additionally, we assess the robustness of our measurements and compare the time-scales determined here to those of previous studies of molecular cloud lifetimes, both in the LMC and in other nearby galaxies. We conclude with a top-level discussion of the condensation of molecular clouds from the diffuse interstellar medium and the destruction of molecular clouds by feedback processes.

\subsection{The formation of molecular clouds}

\label{mc_formation}

\begin{figure*}
	\includegraphics[width=0.49\linewidth]{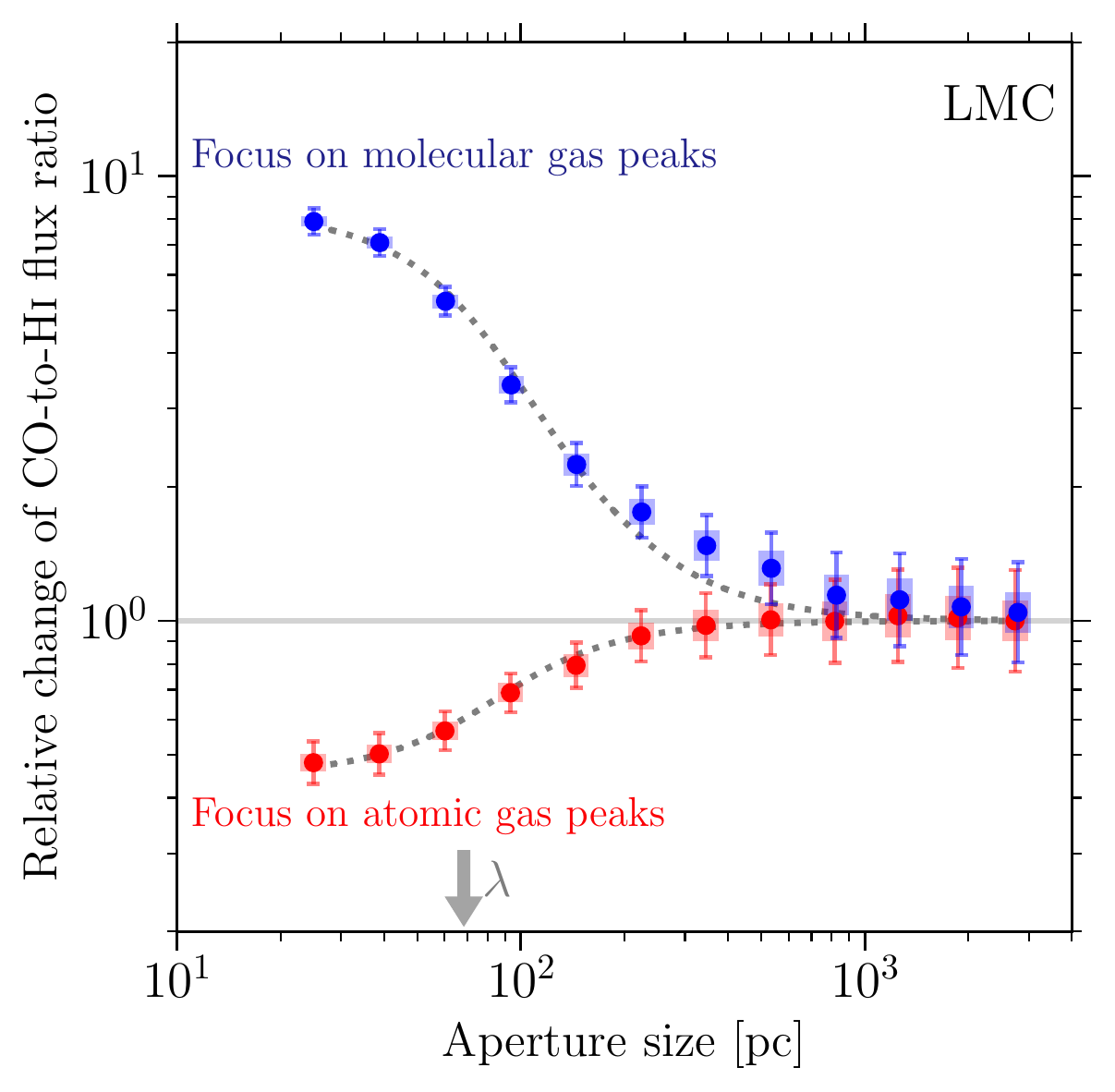}
	\includegraphics[width=0.49\linewidth]{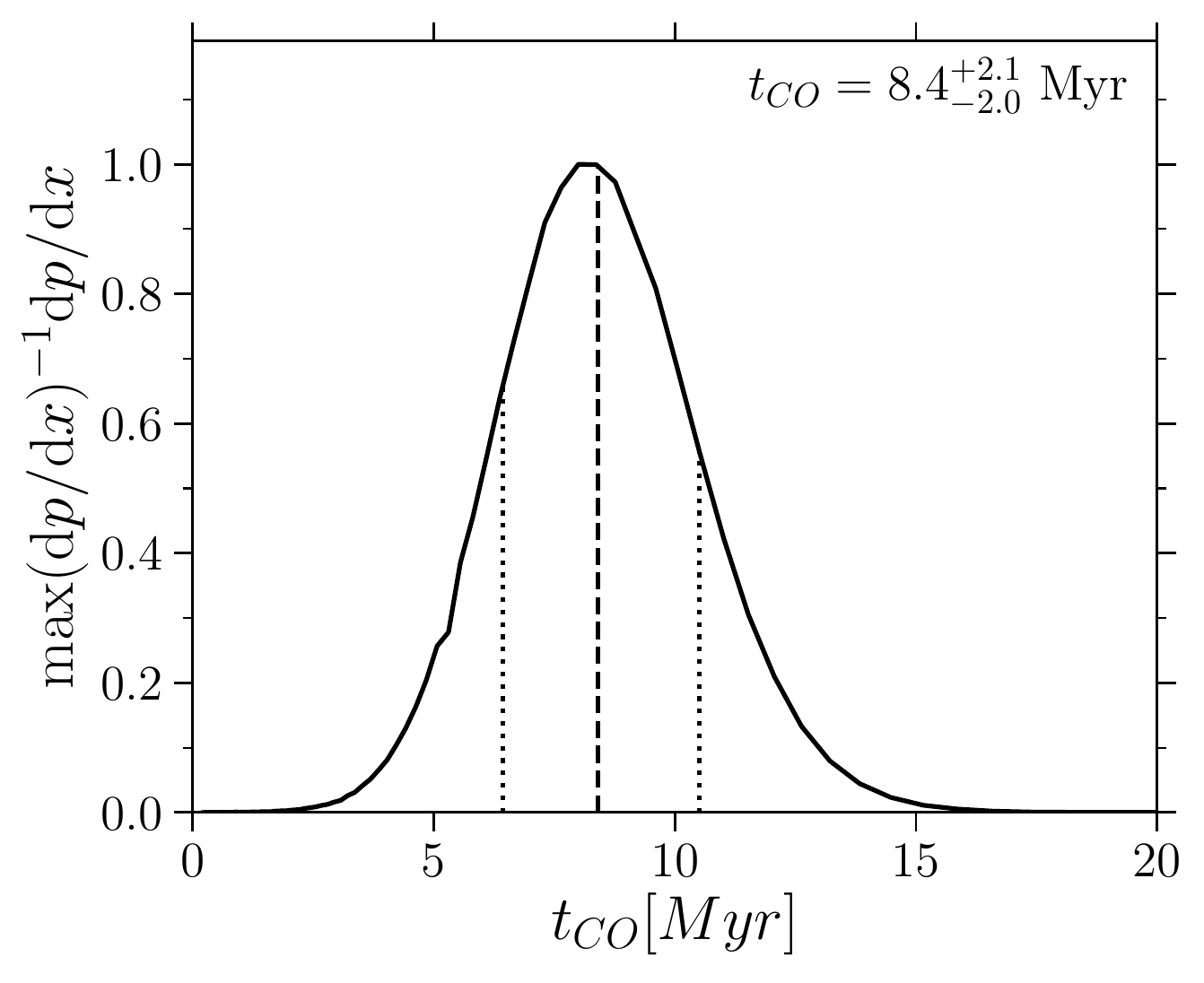}
	\includegraphics[width=0.49\linewidth]{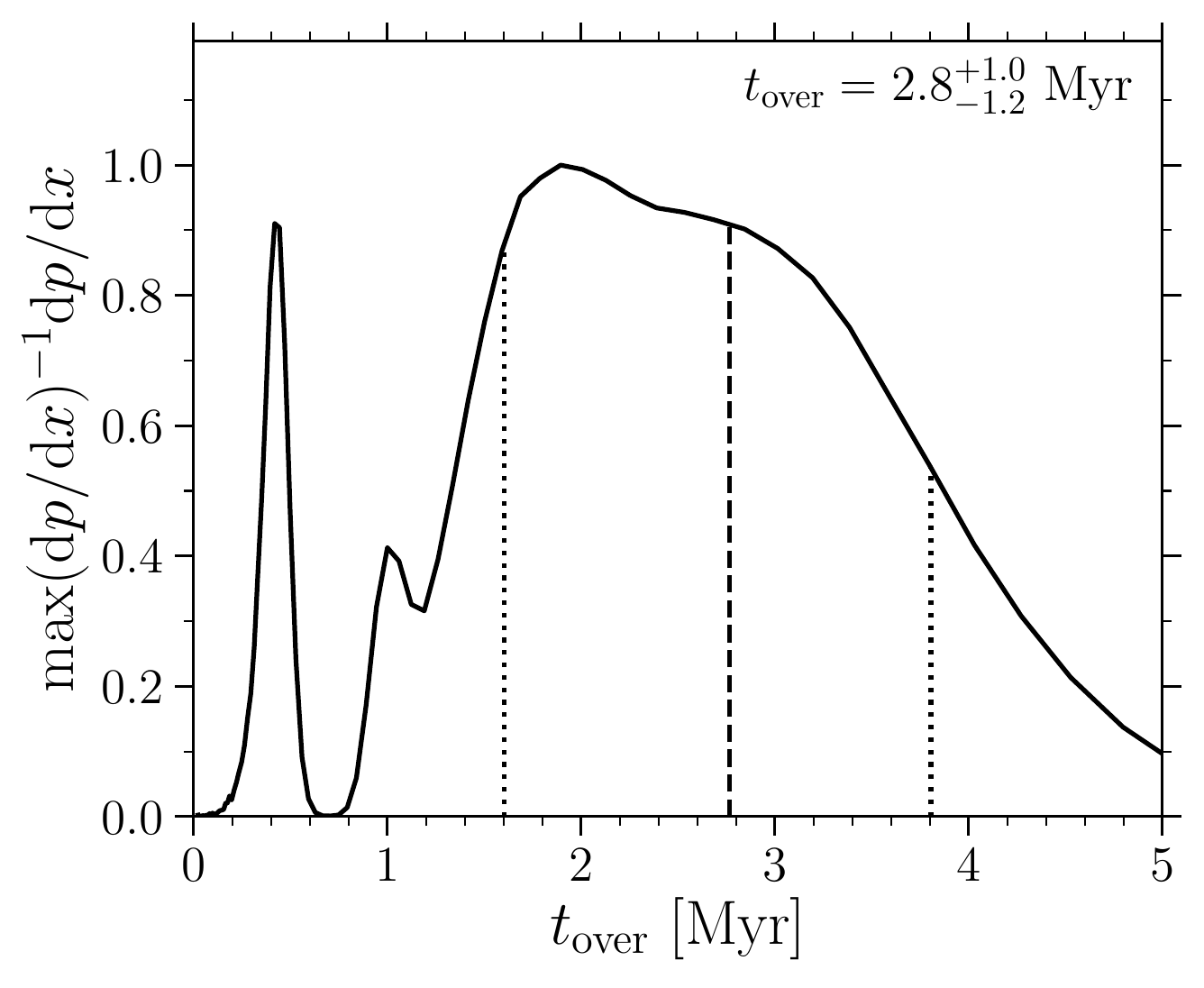}
	\includegraphics[width=0.49\linewidth]{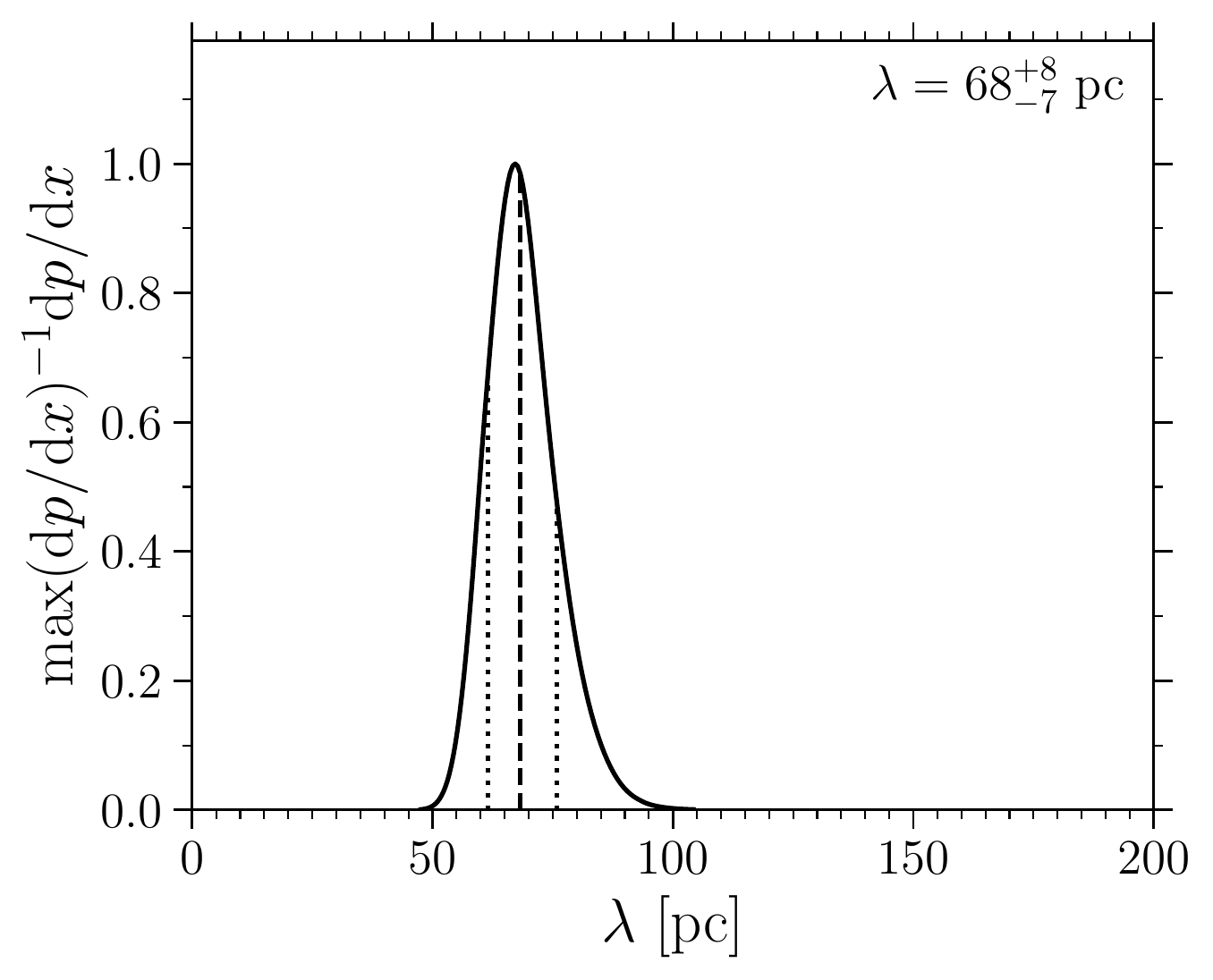}
	\caption{\label{HICOfigure} Results of running the {\sc Heisenberg} code using the H{\sc i} image of the LMC as a reference map and the MAGMA CO moment~0 image as the target map. Upper-left: the CO-to-H{\sc i} emission flux ratio as a function of aperture size. Upper-right: the resulting PDF of the molecular cloud lifetime ($t_{\text{CO}}$). Lower-left: the PDF of the time-scale for which H{\sc i} and CO emission co-exist. Lower-right: the PDF of the characteristic separation length scale ($\lambda$) between atomic and molecular clouds in the LMC. Note that while the precise time-scale for which the tracers co-exist is poorly constrained, it is clear the H{\sc i} emission is present for less than half of the total lifetime of a molecular cloud.} 
\end{figure*}

In much the same way that the circumstances surrounding the death of molecular clouds can be obtained through the time-scale for which massive stars and molecular clouds co-exist, we can also learn about the birth of molecular clouds from the time-scales associated with their atomic cloud progenitors and the overlap time-scale between molecular and atomic clouds. To this end, we repeat our earlier analysis on CO and H$\alpha$, but this time replace the latter with H{\sc i} emission, where as a reference time-scale we adopt the H{\sc i} cloud lifetime derived by \citetalias{Ward2020_HI}, i.e.~$t_{\text{H{\sc i}}} = 48$\,Myr. In Figure~\ref{HICOfigure}, we present the relative change of the CO-to-H{\sc i} flux ratio as a function of spatial scale, as well as the one-dimensional PDFs for each of the three fitted parameters. All quantities are listed in the final column of Table~\ref{fun_tab}.

From this experiment, we derive a time-scale for CO emission in the LMC of 8.4$\substack{+2.1\\-2.0}$\,Myr, which is qualitatively consistent with that derived using H$\alpha$ emission as a reference. However, when experiments with different reference time-scales are performed, the one with the most symmetric de-correlation diagram (reflecting the tracer pair with the most similar visibility time-scales) is the most robust \citep{Kruijssen2018}. In this context, this means that the molecular cloud lifetime estimated using H$\alpha$ emission is the most reliable in a quantitative sense. We also derive a characteristic separation length $\lambda = 68\substack{+8\\-7}$\,pc, considerably shorter than that calculated for CO and H$\alpha$ as well as shorter than that derived in \citetalias{Ward2020_HI}. 

Most importantly, we infer that H{\sc i} and CO emission co-exist for $t_{\text{over}_{\text{H{\sc i},CO}}} = 2.8\substack{+1.0\\-1.2}$\,Myr. This is significantly shorter than the derived molecular cloud lifetime (see Section~\ref{fundamentalParameters}) and suggests that the cloud-wide conversion from atomic to molecular gas is relatively rapid. This is in agreement with the small cold H{\sc i}/CO ratio measured towards molecular clouds using H{\sc i} narrow self-absorption in the LMC \citep{Liu2019}, showing that molecular clouds are constituted of more than 99~per~cent of molecular gas. Similar results are also observed for molecular clouds in the Milky Way \citep{Li2003, Krco2010}.

In combination with the results of our analysis using the H$\alpha$ and H{\sc i}, obtained in \citetalias{Ward2020_HI}, we establish that there is an isolated molecular phase in which the CO emitting clouds are not associated with a significant atomic gas overdensity, prior to the emergence of unembedded H\,{\sc ii} regions. In \citetalias{Ward2020_HI}, we speculated that such an isolated molecular phase was a possibility and this is now confirmed by the current work. Note that by isolated, we mean only with respect to atomic H{\sc i} emission and H$\alpha$ emission, i.e.~a molecular cloud without unembedded massive stars or a significant H{\sc i}-emitting gas reservoir. Here, we draw no conclusions regarding the abundance of H{\sc i}-dark atomic gas or the presence of embedded star formation events. This latter possibility has been explored in \cite{Kim2021}, by using the 24$\mu$m emission as a tracer of embedded star formation. They find that the deeply obscured phase of star formation (during which 24$\mu$m emission is visible, but not H$\alpha$) lasts for 3.8\,Myr in the LMC. 

\subsection{Comparison with model time-scale predictions}

We now compare the molecular cloud lifetimes derived in this work with the predicted time-scales of \citet{Jeffreson2018}, based on galactic dynamics, as well as the analytic predictions of the cloud-scale free-fall time and crossing time. In \citetalias{Ward2020_HI}, we predicted the expected atomic cloud lifetime in the LMC under the influence of galactic dynamical processes, finding that this predicted lifetime is 50\,Myr, consistent with the measured atomic gas cloud time-scale of $48^{+13}_{-8}$\,Myr. This predicted atomic cloud lifetime is obtained using the observed HI column density map. In the current work, we aim to predict and measure the \textquoteleft total gas\textquoteright\ cloud life-time (although this does not include CO-dark molecular gas or optically thick H{\sc i}). The prediction is obtained in the same way as in \citetalias{Ward2020_HI}, but this time using the total gas mass surface density map, combining HI and CO emission. The addition of the molecular gas mass decreases the mid-plane free-fall time, resulting in a predicted lifetime for the total gas condensation of 44\,Myr. This is still significantly longer than the measured molecular cloud lifetime in the LMC of $11.8^{+2.7}_{-2.2}$\,Myr, and of a magnitude similar to the measured total gas cloud lifetime of 55-60\,Myr, which combines the durations of the atomic and molecular cloud lifetimes.

As in \citet{Kruijssen2019} and \citet{Chevance2020}, we define the molecular cloud free-fall time-scale as:
\begin{equation}
    t_{\text{ff,cl}} = \sqrt{\frac{\pi^{2}r^{3}_{\text{GMC}}}{10GM_{\text{GMC}}}} \text{,}
\end{equation}
where $r^{3}_{\text{GMC}}=1.9r_{\text{CO}}$ \citep{Kruijssen2019}. The molecular cloud mass is given by $M_{\text{GMC}} = \mathcal{E}_{\text{H}_{2}}\Sigma_{H_{2}}\pi(\lambda/2)^{2}$, where $\mathcal{E}_{\text{H}_{2}}$ is the surface density contrast on the size scale $\lambda$ relative to the surface density measured across the entire field of view, $\Sigma_{\text{H}_{2}}$ \citep{Kruijssen2018}. We elect to make use of the resulting values from the {\sc Heisenberg} code for both self-consistency between the measured lifetimes and predicted time-scales, and also for consistency between this work and the multiple-galaxy study of \citet{Chevance2020}. For the LMC out to a radius of 3.3\,kpc, we derive an average cloud-scale free-fall time-scale of 11.5\,Myr. This is consistent with the measured molecular cloud lifetime in the LMC.

The crossing time is defined as
\begin{equation}
    t_{\text{cr}} = \frac{r_{\text{GMC}}}{\sigma_{\text{vel}}}\text{,}
\end{equation}
where $\sigma_{\text{vel}}$ is the one-dimensional velocity dispersion. In this case we use the one-dimensional velocity dispersion map from the MAGMA survey. Pixels in this velocity dispersion map are masked where no flux is measured for the corresponding pixel in the filtered moment~0 map. This velocity dispersion map therefore only includes regions where the compact flux component is detected. Using the CO peak radius derived with the {\sc Heisenberg} code and the mean velocity dispersion measured across the galaxy, we calculate a mean cloud crossing time of 21.4$\substack{+1.0\\-0.7}$\,Myr. While significantly shorter than the time-scale derived based on galactic dynamics, this time-scale is longer than both the predicted free-fall time-scale as well as the measured lifetime of molecular clouds in the LMC. The ratio between the free-fall time and the crossing time implies a typical molecular cloud virial parameter of $\alpha_{\rm vir}\approx0.85$, consistent with virial balance between the kinetic and potential energy terms.

\begin{figure*}
    \centering
    \includegraphics[width=0.9\linewidth]{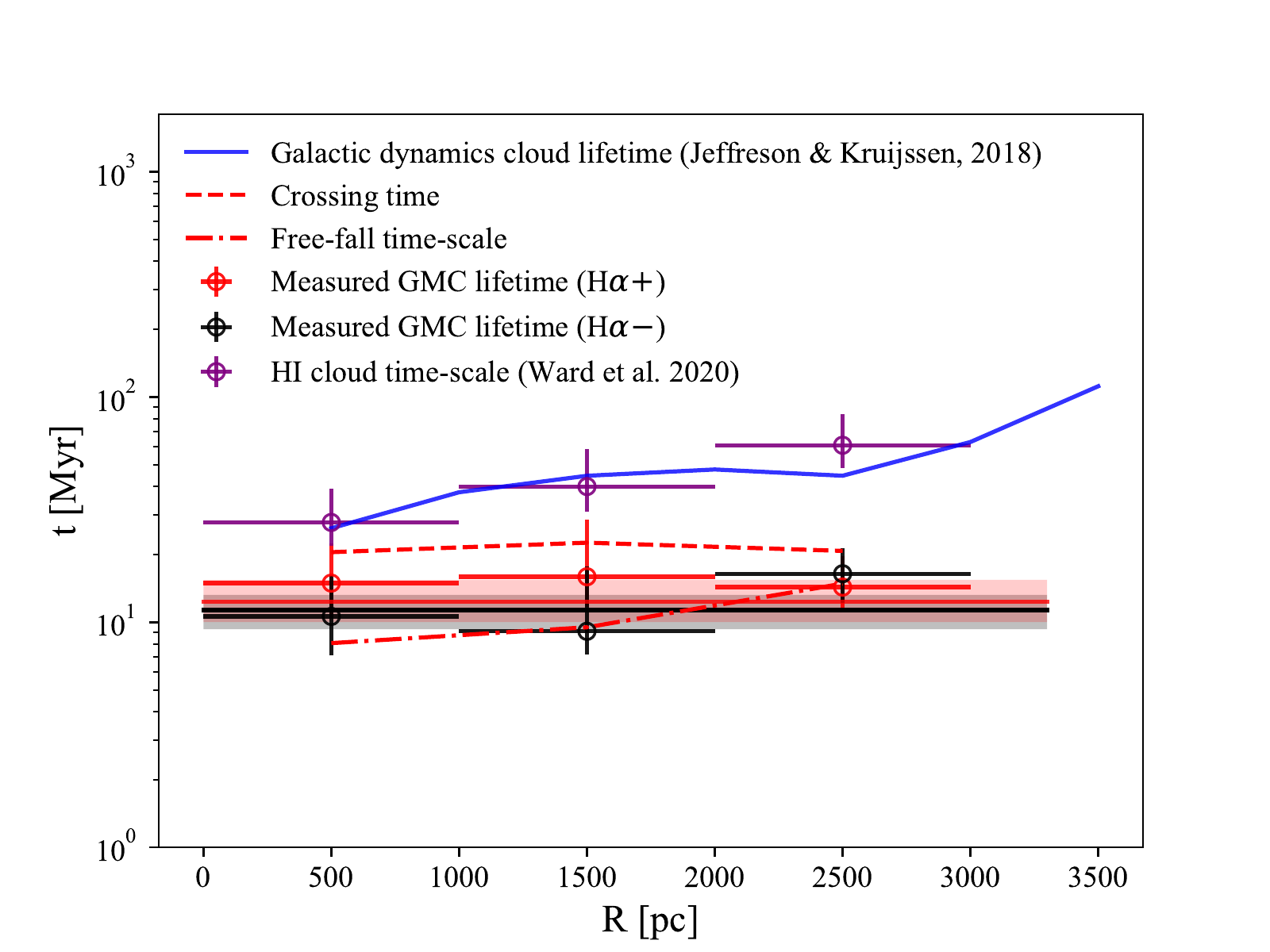}
    \caption{\label{predictCO} Cloud lifetimes as a function of galactocentric radius. The measured molecular cloud lifetimes measured using the continuum-substracted SHASSA H$\alpha$ and MCELS H$\alpha$ images across three radial bins as reference maps are shown in black and red, respectively. The solid red and black lines represent the measured average time-scale across the entire field of view, with the corresponding uncertainties denoted by the shaded areas surround the average lines. The predicted cloud lifetime based on galactic dynamics is shown as a function of radius by the solid clue curve. The cloud-scale free-fall time-scale and crossing time predicted for each bin are denoted by the dash-dot and dashed lines, respectively. Across the entirety of the LMC, the measured molecular cloud lifetimes are inconsistent with those predicted based on galactic dynamics, which provide a close match to the atomic cloud lifetime. By contrast, the time-scales based solely on internal processes of the molecular clouds (the crossing and free-fall time-scales) fall much closer to the measured molecular cloud lifetimes.}
\end{figure*}

In Figure~\ref{predictCO}, we present the measured molecular cloud lifetime as a function of galactocentric radius, measured in three radial bins that are 1~kpc in width. This is done using both the MCELS H$\alpha$ plus continuum emission map (red) and the SHASSA continuum subtracted H$\alpha$ map (black). Also shown are the galactic average time-scales as presented in Table~\ref{fun_tab}. The cloud lifetimes measured within each radial bin are statistically consistent with the average time-scales at all galactocentric radii, indicating that the time-scales associated with molecular clouds do not vary strongly as a function of galactocentric radius in the LMC. The predicted molecular cloud lifetime based on galactic dynamical processes such as spiral arm crossings, epicyclic perturbations, shear, and cloud-cloud collisions \citep{Jeffreson2018} is shown as a function of radius by the solid blue curve. At no radius is the measured cloud lifetime consistent with that predicted from galactic dynamics. It is therefore clear that galactic dynamical processes cannot control the molecular cloud lifetime based on the time-scales measured in this work.

Instead, the measured molecular cloud lifetimes are far more consistent with time-scales associated with internal cloud processes, i.e.~the molecular cloud free-fall time-scale and the crossing time. This suggests that the molecular clouds in the LMC are decoupled from galactic dynamics, in stark contrast with the time-scales associated with atomic gas clouds, for which galactic dynamics can predict the cloud lifetimes at all radii (\citetalias{Ward2020_HI}). Given the average molecular gas surface density of the LMC ($\Sigma_{\rm H_2} \sim 2$\,M$_\odot$\,pc$^{-2}$; \citealt{Jameson2016,Schruba2019}), this result is in agreement with the finding of \citet{Chevance2020}, who show that in environments of low kpc-scale molecular gas surface densities ($\Sigma_{\rm H_2} \lesssim 8$\,M$_\odot$\,pc$^{-2}$), cloud lifetimes are governed by internal dynamical processes (with characteristic time-scales close to the free-fall time and the crossing time), while galactic dynamics dominates at higher surface densities.

\subsection{Robustness of results}

In \citet{Kruijssen2018}, a series of guidelines are established that should be followed in order to ensure a robust application of the {\sc Heisenberg} code used in this work. These guidelines are the result of extensive testing of the methodology across a broad parameter space. As in \citetalias{Ward2020_HI}, we address each of these criteria for the application of the statistical method to the CO and H$\alpha$ maps:

\begin{enumerate}
    \item The duration of the time-scales associated with the two tracers in question must not differ by more than an order of magnitude. Indeed, the derived molecular cloud lifetime does not differ from the H$\alpha$ reference time-scales by more than a factor of 2.
    \item The characteristic length scale satisfies the criterion $\lambda \geq N_{\text{res}} l_{\text{ap,min}}$, where $l_{\text{ap,min}}$ is the minimum aperture size within which the flux ratio is measured (the spatial resolution of the observations imposes a lower limit to this value), where $N_{\text{res}}=\{1,1.5\}$ in order to accurately determine $\{t_{\text{CO}},t_{\text{fb}}\}$. 
    \item The number of emission peaks in each tracer map must be greater than 35 in order to achieve a precision of 0.2\,dex. Higher levels of precision can be achieved when identifying greater numbers of emission peaks. This number of emission peaks is exceeded in each application of {\sc Heisenberg} in this study, including those that are applied to radial bins.
    \item By focusing an aperture on an emission peak in a particular tracer, there should not be a deficit of this tracer relative to the galactic-scale average. This condition is satisfied after filtering the diffuse emission as specified in Section~\ref{uncertaintyPrinciple}. 
    \item The star formation rate across the field of view considered should not change by more than 0.2~dex over the time spanned by the typical region lifecycle. Based on the star formation rate history determined by \citet{Harris2009}, and because we mask 30~Doradus, which is likely the region to exhibit the largest change in star formation rate recently, we consider that the star formation rate in the LMC has indeed been sufficiently constant over the past $\sim 20$\,Myr.
\end{enumerate}

Having satisfied all of the above criteria, we establish that the time-scales and subsequently derived parameters presented in this work are robust within our methodological framework. In the next section we discuss how the time-scales measured in this study compare to previous estimates of the molecular cloud lifetime in the LMC.

\subsection{Comparison with previous studies of the LMC}

\label{comparisonSection}

\citet{Kawamura2009} estimate a range of 20-30\,Myr for the molecular cloud lifetime in the LMC. This is significantly longer than the average CO emission time-scale measured in this study of 11.8$\substack{+2.7\\-2.2}$\,Myr, because our measurement falls below the lower bound of the range from \citet{Kawamura2009} by approximately $3\sigma$. This warrants further discussion, because the methodology employed in our analysis has been demonstrated to yield robust measurements across a wide variety of test cases using simulated galaxies \citep{Kruijssen2018, Fujimoto2019, Haydon2020b, Jeffreson2021, Semenov2021} and observations \citep{Kruijssen2019,Chevance2020,Kim2021}. What is the source of the disagreement between our results and those of \citet{Kawamura2009}?

The molecular cloud lifetime determined by \citet{Kawamura2009} implies a time-scale for which H\,{\sc ii} regions and molecular clouds co-exist of 20\,Myr. This is significantly longer than the average time-scale for which H\,{\sc ii} regions are visible in H$\alpha$ ($\lesssim 10$\,Myr; e.g.\ \citealt{Leroy2012, Haydon2020}). This long overlap time-scale would imply an average feedback velocity of $\sim$1\,km\,s$^{-1}$ (based on the cloud sizes reported by \citealt{Fukui2008}), inconsistent with observed expansion velocities of H\,{\sc ii} regions in the LMC. In contrast, the overlap time-scales determined in this work correspond to feedback velocities velocities of $\sim$12\,km\,s$^{-1}$, consistent with observed velocities of early-stage feedback mechanisms in the LMC (see Section~\ref{feedbackVelocity}). This suggests that the durations of the various phases of the evolutionary timeline from \citet{Kawamura2009} may have been overestimated.

There are a number of reasons why the time-scales associated with CO-emitting clouds in the LMC differs between our work and that of \citet{Kawamura2009}, but we focus on the most important one here.\footnote{Other potential reasons for disagreement are differences in accounting for chance projections (which is self-consistent in our analysis), the assumption that all SWB0 clusters are formed in clouds designated as `GMCs' by \citet{Kawamura2009}, and the assumption that these clusters are representative for all clusters and associations up to the age of 10\,Myr.} The classification used by \citet{Kawamura2009} to construct the evolutionary timeline is based on subjective choices that may have quantitatively affected the result. The three phases adopted by \citet{Kawamura2009} are `type I molecular clouds' containing no signs of star formation,`type II molecular clouds' containing H\,{\sc ii} regions, and `type III molecular clouds' containing both H\,{\sc ii} regions and `SWB0' stellar clusters with ages $<$10\,Myr. Like our method, the cloud-counting approach of \citet{Kawamura2009} requires the use of a `reference time-scale' to which the rest of the evolutionary timeline is anchored. \citet{Kawamura2009} choose to anchor their timeline to the lifetime of type III molecular clouds, which is taken to be 6.6\,Myr, because 66\% of the `SWB0' stellar clusters with ages $<$10\,Myr are found to be associated with molecular clouds. Based on the relative number counts, they then obtain lifetimes for the type I and type II molecular cloud phases of 6~and 13\,Myr, respectively, and a total cloud lifetime of 26\,Myr.

The difference relative to our work arises mainly due to the choice made by \citet{Kawamura2009} to treat type III molecular clouds as an evolutionary phase separate to the type II molecular clouds that only contain H\,{\sc ii} regions. However, the presence of H\,{\sc ii} regions requires the presence of (the progenitors of) young stellar clusters or associations that have ages $<$10\,Myr. We therefore believe that the 10\,Myr reference time-scale used by \citet{Kawamura2009} should apply to the combination of both the type II and type III molecular cloud phases, as well as the SWB0 clusters unassociated with molecular clouds. When assigning the 10\,Myr age bin to the combination of type II/III molecular clouds and isolated SWB0 clusters from \citet{Kawamura2009}, their total molecular cloud lifetime becomes 11.3\,Myr, which is consistent with the results of this work. We caution that this consistency does not necessarily indicate that the cloud-counting methodology is reliable, but rather highlights its extreme sensitivity to subjective choices in classification.

\subsection{Molecular cloud time-scale in context}

In \citetalias{Ward2020_HI}, we presented a first empirical measurement of the atomic cloud lifetime in the LMC, deriving a time-scale of $t_{\text{H{\sc i}}} = 48\substack{+13\\-8}$\,Myr. This is almost five times longer than the molecular cloud lifetime measured here. The fact that the subsequent higher-density molecular phase is shorter-lived than atomic clouds suggests that the early stage of star formation, ranging from the condensation of gas clouds from the diffuse interstellar medium to the onset of star formation, is an accelerating process. We also find that the time-scale for which H{\sc i} and H$\alpha$ emission co-exist is consistent with zero, with an upper limit of $t_{\text{over(H}\alpha\text{,H{\sc i}})}<1.7$\,Myr. This suggests that there may be an isolated molecular phase before the onset of star formation, during which no H{\sc i} emission is present. Indeed, we find an overlap time-scale for H{\sc i} and CO emission of 2.8\,Myr, or just 24~per~cent of the molecular cloud lifetime, confirming that H{\sc i} emission is not present throughout most of the lifetime of molecular clouds. Combining this with the overlap time-scale between H$\alpha$ emission and CO emission, $t_{\text{fb}}=1.2\substack{+0.3\\-0.2}$\,Myr, we find that there is indeed a phase (lasting for $7.8$\,Myr) during which no H{\sc i} emission is present prior to the emergence of H\,{\sc ii} regions.

The measured CO-emitting cloud lifetime in the LMC ($t_{\text{CO}}=$11.8$\substack{+2.7\\-2.2}$\,Myr) is remarkably similar to the average GMC lifetime measured in NGC~300 of $t_{\text{CO}}=10.8\substack{+2.1\\-1.7}$\,Myr \citep{Kruijssen2019}. \citet{Chevance2020} present molecular cloud lifetime measurements for nine nearby spiral galaxies. All but one of these exhibit time-scales that are longer than that measured in the LMC. NGC~5068 exhibits a GMC lifetime of 9.6$\substack{+2.9\\-1.8}$\,Myr, consistent within uncertainties with that determined for the LMC. With a stellar mass of 2.3$\times$10$^{9}$\,$M_{\sun}$, NGC~5068 is the lowest-mass galaxy in the sample of \citet{Chevance2020} and has a similar mass to that of the LMC. Together with NGC~300 and NGC~5068, the LMC lies in a low-mass, low-metallicity region of the parameter space spanned by existing studies of molecular cloud lifetimes, and exhibits one of the shortest measured molecular cloud lifetimes. Across a sample of 54 nearby galaxies, \citet{Kim2022} also find a strong trend of increasing cloud lifetime with galaxy mass, consistent with the results reported here.

\citet{Chevance2020} show that molecular cloud lifetimes decouple from galactic-dynamical time-scales below a kpc-scale molecular gas surface density of 8\,M$_{\text{sun}}$\,pc$^{-2}$. With an average molecular gas density of $\sim2$\,M$_{\text{sun}}$\,pc$^{-2}$, the LMC lies comfortably within the low-density density regime proposed by \citet{Chevance2020}. Indeed, we show here that even though galactic dynamics set the atomic cloud lifetime in the LMC (\citetalias{Ward2020_HI}), most molecular clouds in the LMC are decoupled from the effects of galactic dynamics and are instead governed by internal processes (although exceptions like 30~Doradus may still exist\footnote{Another such example is the Headlight Cloud in NGC628, which rapidly accumulates gas due to its location at a resonance \citep{Herrera2020}, whereas most clouds in NGC628 are expected to be governed by internal dynamics \citep{Chevance2020} like in the LMC.}).

\section{Conclusions}

\label{conclusions}

In this paper, we have presented empirically-derived time-scales for the lifecycle of molecular clouds in the LMC, from their condensation from the diffuse ISM to their destruction by stellar feedback. 

\begin{enumerate}
	\item The average molecular cloud lifetime in the LMC is 11.8$\substack{+2.7\\-2.2}$\,Myr. This is approximately a factor of two (or $3\sigma$) shorter than the lower bound of the previously estimated cloud lifetime in the LMC based on counting the numbers of objects classified as molecular clouds or young stellar populations \citep{Kawamura2009}.
	\item The relatively short molecular cloud time-scale is consistent with internal processes such as gravitational free-fall or the cloud crossing time rather than galactic dynamical time-scales such as shear, cloud-cloud collisions, or spiral arm passages. This means that the lifetime of molecular clouds is decoupled from galactic dynamics and regulated by internal dynamics, in clear contrast to the lower-density population of atomic clouds in the LMC.
	\item The overlap time between molecular clouds and H\,{\sc ii} regions, representing the duration of the feedback phase during which the cloud is dispersed by young massive stars, is found to be 1.2$\substack{+0.3\\-0.2}$\,Myr. In combination with the derived average radius of the molecular clouds in this study ($\sim 14$\,pc), this implies a typical feedback velocity of $\sim 12 \pm$2\,km\,s$^{-1}$, consistent with the observed expansion velocities of H\,{\sc ii} regions in the LMC.
	\item The characteristic separation length between star forming regions is measured to be 92\,pc. This is consistent with the estimated molecular gas scale height of the LMC.
	\item In combination with the results of our previous paper on the atomic cloud lifetime, we establish a time-scale for the formation of molecular clouds. The total time-scale for which molecular clouds can be considered to be in the process of forming (i.e.~from the initial emergence of atomic gas clouds) is 48\,Myr, as established in the previous paper. The time-scale for which molecular clouds co-exist with atomic gas clouds after CO is first detected is short, at 2.8$\pm$1\,Myr.
	\item The molecular cloud lifetime reported in this paper is shorter than the result previously found by \citet{Kawamura2009}, who obtained $20{-}30$\,Myr by comparing the number counts of clouds, H\,{\sc ii} regions, and young stellar clusters and associations with ages $<10$\,Myr. We argue that this difference results from the assumption made by \citet{Kawamura2009} that the 10 Myr age bin used to calibrate the time-scale measurement should exclude clouds that only contain H\,{\sc ii} regions and do not contain any young stellar clusters. When extending this age bin to also include the clouds that only contain H\,{\sc ii} regions, the cloud lifetime obtained through the method of \citet{Kawamura2009} becomes 11.3\,Myr, which is consistent with our measurement. The extreme sensitivity of the cloud-counting methodology to these types of subjective choices acts as a further motivation to instead adopt flux-based measurements as in the present paper.
\end{enumerate}

In this work, we establish that molecular clouds in the LMC live for 11.8$\substack{+2.7\\-2.2}$\,Myr. The derived time-scale is not dependent on subjective methods for classifying clouds or young stellar regions. These results sketch a molecular cloud lifecycle in the LMC in which molecular clouds condense from the atomic interstellar medium in an accelerating process resulting in a near-complete phase change from atomic to molecular gas before the onset of massive star formation. Once formed, molecular clouds are largely decoupled from galactic-scale dynamical processes, with lifetimes consistent with internal dynamics. The mechanical dispersal of molecular clouds occurs on a time-scale shorter than expected for the onset of the first supernovae, which implies feedback dispersal velocities that are similar to the observed expansion velocities of H\,{\sc ii} regions in the LMC. We therefore conclude that early feedback processes such as photoionisation are likely to drive molecular cloud destruction in the LMC.

\section*{Acknowledgements}
We thank an anonymous referee for helpful suggestions.
JLW, JMDK, and JK gratefully acknowledge support from the Deutsche Forschungsgemeinschaft (DFG, German Research Foundation) -- Project-ID 138713538 -- SFB 881 (\textquotedblleft The Milky Way System\textquotedblright, subproject B2).
JMDK, MC, and JK gratefully acknowledge funding from the DFG through the DFG Sachbeihilfe (grant number KR4801/2-1).
JMDK and MC gratefully acknowledge funding from the DFG through an Emmy Noether Research Group (grant number KR4801/1-1), as well as from the European Research Council (ERC) under the European Union's Horizon 2020 research and innovation programme via the ERC Starting Grant MUSTANG (grant agreement number 714907).
MC gratefully acknowledges funding from the DFG through an Emmy Noether Research Group (grant number CH2137/1-1).

\section*{Data availability}

The data underlying this article are available upon reasonable request to the corresponding author.




\bibliographystyle{mnras}
\bibliography{bibliography} 




\appendix

\section{Combining time-scale measurements}

\label{combiningPDFs}

In Section~\ref{results}, we present time-scales for CO emission that are calculated by combining the results of individual {\sc Heisenberg} runs. The simplest way of making this combination is to use the average weighted by the uncertainties in the derived time-scale. However, this can be unreliable if the estimates are correlated. Therefore, we adopt the method of \citet{Lyons1988} and combine the derived PDFs of $t_{\rm CO}$ using a set of weights $\alpha$ such that the combined PDF $p(x)$ is defined as
\begin{equation}
\label{eqn1_combPDF}
p(x) = \sum_{i} \alpha_{i}p_{i}(x) ,
\end{equation}
where $p_i(x)$ is the PDF of the observable of interest from an individual experiment $i$. The values of the coefficients $\alpha_i$ are chosen to minimise the total variance, $\sigma^2$, while obeying the condition that $\sum_{i}\alpha_i=1$. The total variance is defined as the sum over the error matrix $\textbf{E}$ times the products of the weights $\alpha$:
\begin{equation}
\sigma^{2} = \sum_{i} \sum_{j} E_{ij}\alpha_{i}\alpha_{j} ,
\end{equation}
where the error matrix \textbf{E} is defined as
\begin{equation}
E_{ij} = r_{ij}\sigma_{i}\sigma_{j} ,
\end{equation}
where $r_{ij}$ is the correlation coefficient between two measurements and $\sigma_{i}$ is the standard error implied by the PDF of the measurement $i$ (see below). In the case of combining two time-scales obtained from two different H$\alpha$ maps as in this work, we thus obtain
\begin{equation}
\textbf{E} = \left(
\begin{matrix}
\sigma_{1}^{2} & r_{12}\sigma_{1}\sigma_{2}  \\
r_{21}\sigma_{2}\sigma_{1} & \sigma_{2}^{2}  \\
\end{matrix}
\right) ,
\end{equation}
where the subscripts 1 and 2 correspond to the experiments using the two different star formation tracers, i.e.~H$\alpha -$ and H$\alpha +$.

The degree to which the measured time-scales are correlated between two different star formation tracer maps is defined entirely by the choice of these maps, their respective reference time-scales, and the peak selection in these maps. However, it is not possible a priori to express analytically how the correlation between these two maps propagates into a correlation between the two resulting molecular cloud lifetimes. This complicates the definition of $r_{ij}$.

To determine the correlation coefficient, we adopt the axiom that any deviation from a perfect correlation between the two $t_{\rm CO}$ PDFs must directly reflect a similar deviation between the experiments used to derive the time-scale. We therefore define the correlation coefficient between two $t_{\rm CO}$ measurements, $r_{ij}$, as the median of the difference between the measured and expected ratios between the two PDFs across the range of possible values of $t_{\rm CO}$:
\begin{equation}
r_{ij} = \text{med}\left\{\left[\frac{{\rm d}p_j(t_{\rm CO})}{{\rm d}p_i(t_{\rm CO})}\right]_{\rm meas} -
\left[\frac{{\rm d}p_j(t_{\rm CO})}{{\rm d}p_i(t_{\rm CO})}\right]_{\rm exp}\right\} ,
\end{equation}
where ${\rm d}p_j(t_{\rm CO})/{\rm d}p_i(t_{\rm CO})$ is the derivative between the probabilities for the two PDFs at a variable position $t_{\rm CO}$.

Finally, we must determine the values of $\sigma_i$. The uncertainties on the time-scales derived with {\sc Heisenberg} are asymmetric, which makes this choice non-trivial. However, we only use $\sigma_i$ to determine the weight $\alpha_i$ of each PDF, which weakens its impact on the combined value and uncertainty of $t_{\rm CO}$. For simplicity, we therefore define $\sigma_i$ as the simple geometric average between the upper and lower uncertainties on each value, as in \citet{Lyons1988}.

\begin{figure*}
	\includegraphics[width=0.49\linewidth]{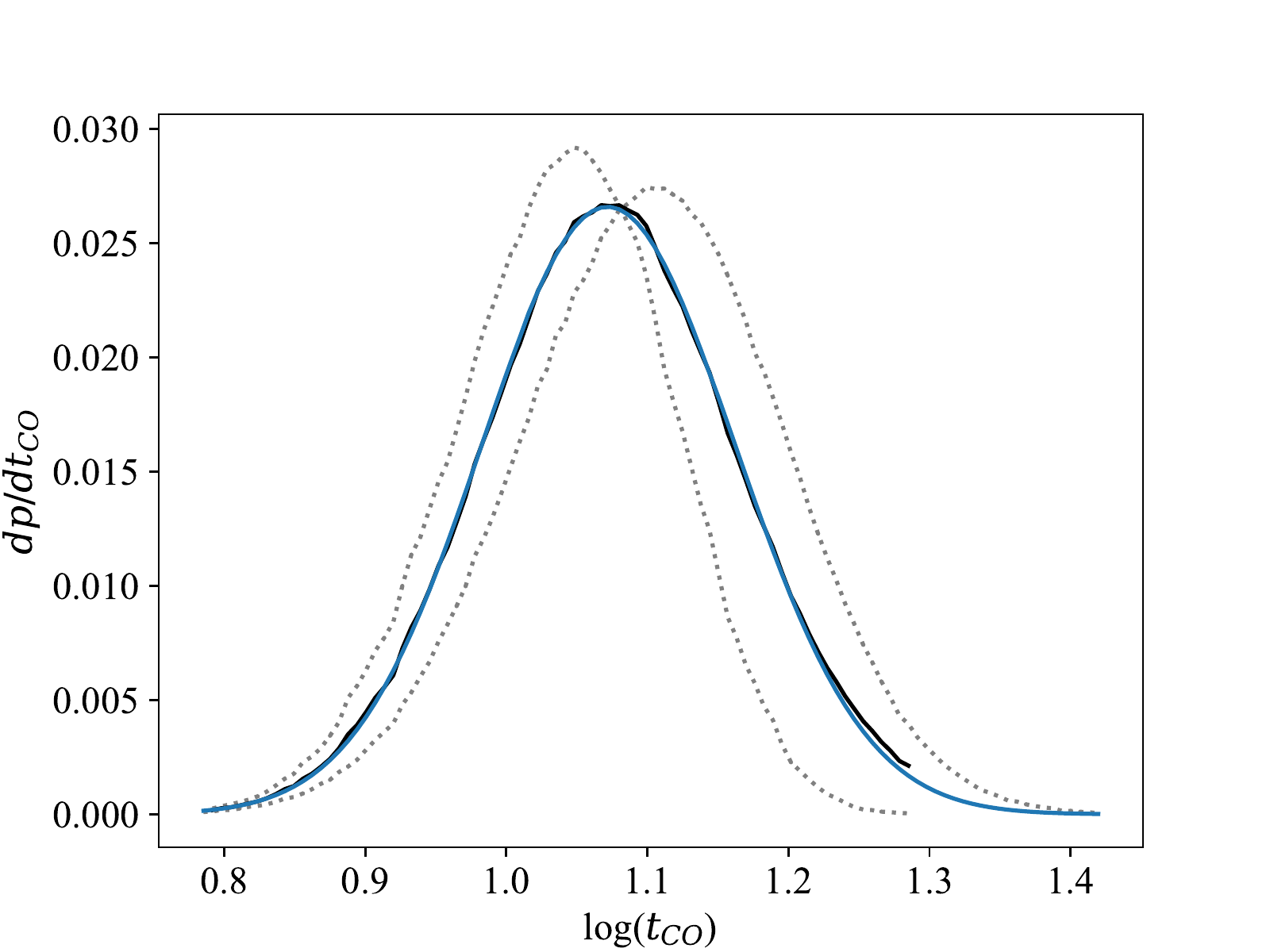}
	\includegraphics[width=0.49\linewidth]{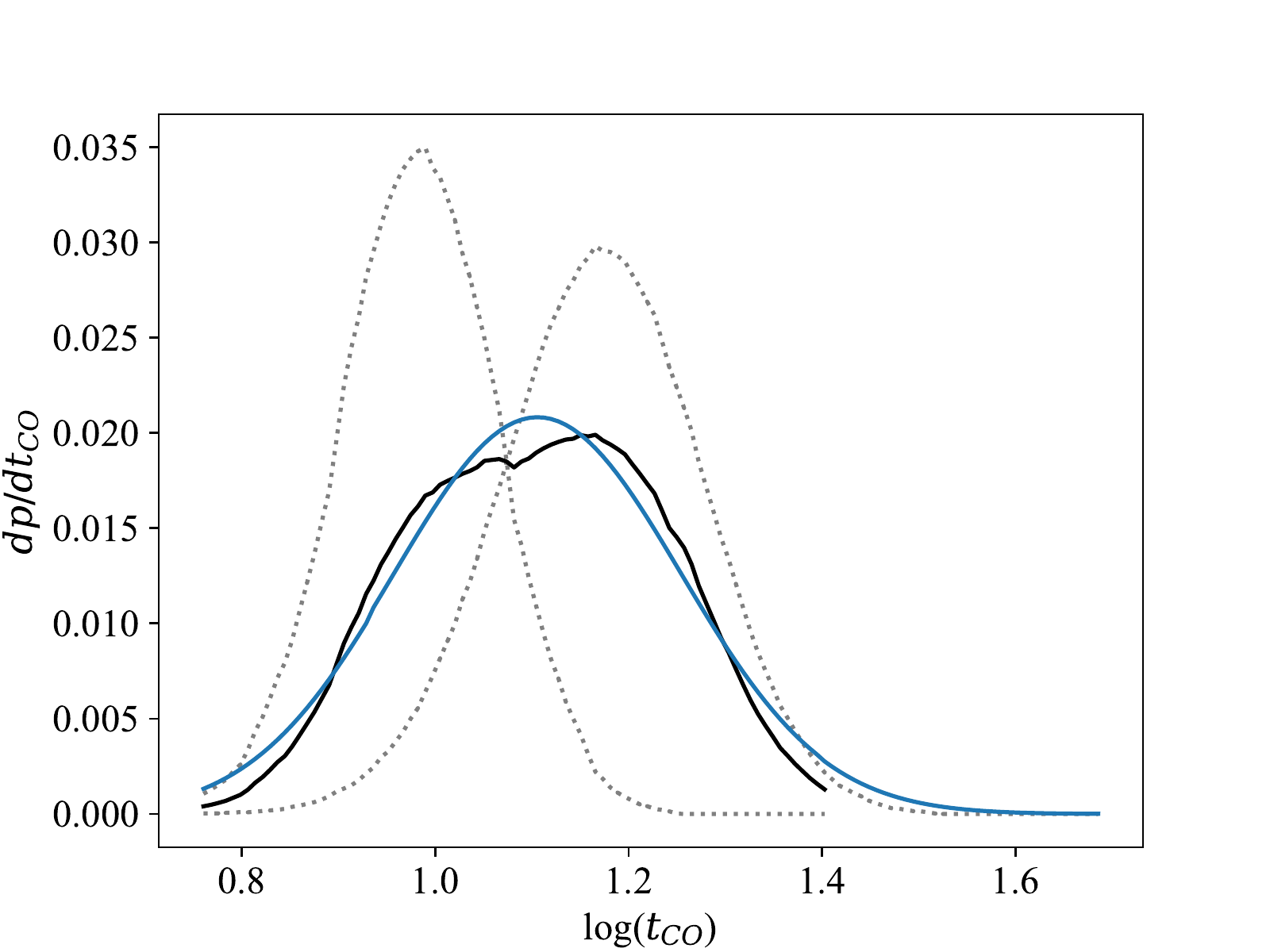}
	\caption{\label{combinedPDFfig} Illustration of the procedure used to combine different experiments performed with {\sc Heisenberg}. In each panel, the two dotted lines show the $t_{\text{CO}}$ PDFs for the runs performed using the two different H$\alpha$ maps. The combined PDF is displayed as a solid black line and the two-sided Gaussian model fitted to that combined PDF is shown in blue. Left: excluding 30~Doradus using a 200\,pc mask. Right: including 30~Doradus.}
\end{figure*}

After calculating the weights $\alpha_i$, the combined PDF is constructed according to equation (\ref{eqn1_combPDF}). To account for the asymmetric nature of the PDF, we fit it with two half Gaussian profiles that connect at the maximum value of the distribution. We show this PDF and the best-fitting two-sided Gaussian models in Figure~\ref{combinedPDFfig}. The combined measurement is then taken to be the mean centroid of the two Gaussian profiles, and the positive and negative uncertainties are defined as the standard deviations of the positive upper and lower fits, respectively.

\section{The effects of limited coverage and sensitivity}

\label{limitedCoverage}

As described in Section~\ref{obsSection}, the MAGMA survey does not provide full coverage of the LMC and indeed, does not observe every molecular cloud reported by \citet{Fukui2008}. In this section, we consider how the limited coverage may impact our results. We perform additional {\sc Heisenberg} experiments to assess the different methods of treating the molecular clouds that are missing from the MAGMA survey. The first experiment uses the MAGMA CO data {\it without} masking the regions around molecular clouds that are only detected in the NANTEN survey of the LMC. In the second experiment, we use the reported positions, sizes, and fluxes of the molecular clouds from the NANTEN survey of the LMC of \citet{Fukui2008}, to model the molecular clouds as two-dimensional Gaussian structures. These model clouds are then inserted into the MAGMA CO map for all clouds that fall outside of the coverage of the MAGMA observations. These experiments are run with both available H$\alpha$ emission maps as reference maps: the MCELS map that includes continuum emission (H$\alpha+$), and the SHASSA maps that is continuum-subtracted (H$\alpha-$).

\begin{figure*}
	\includegraphics[width=0.49\linewidth]{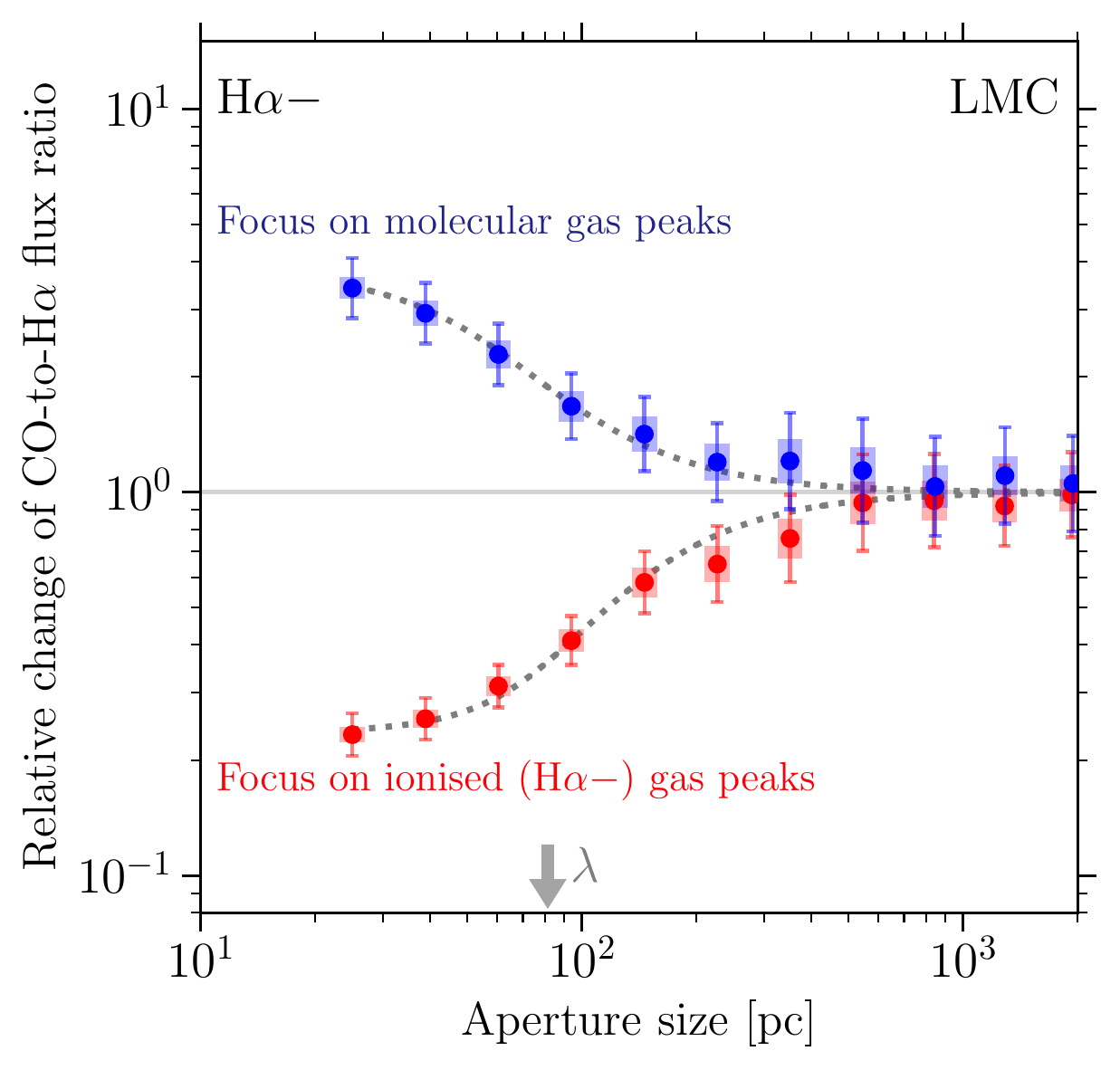}
	\includegraphics[width=0.49\linewidth]{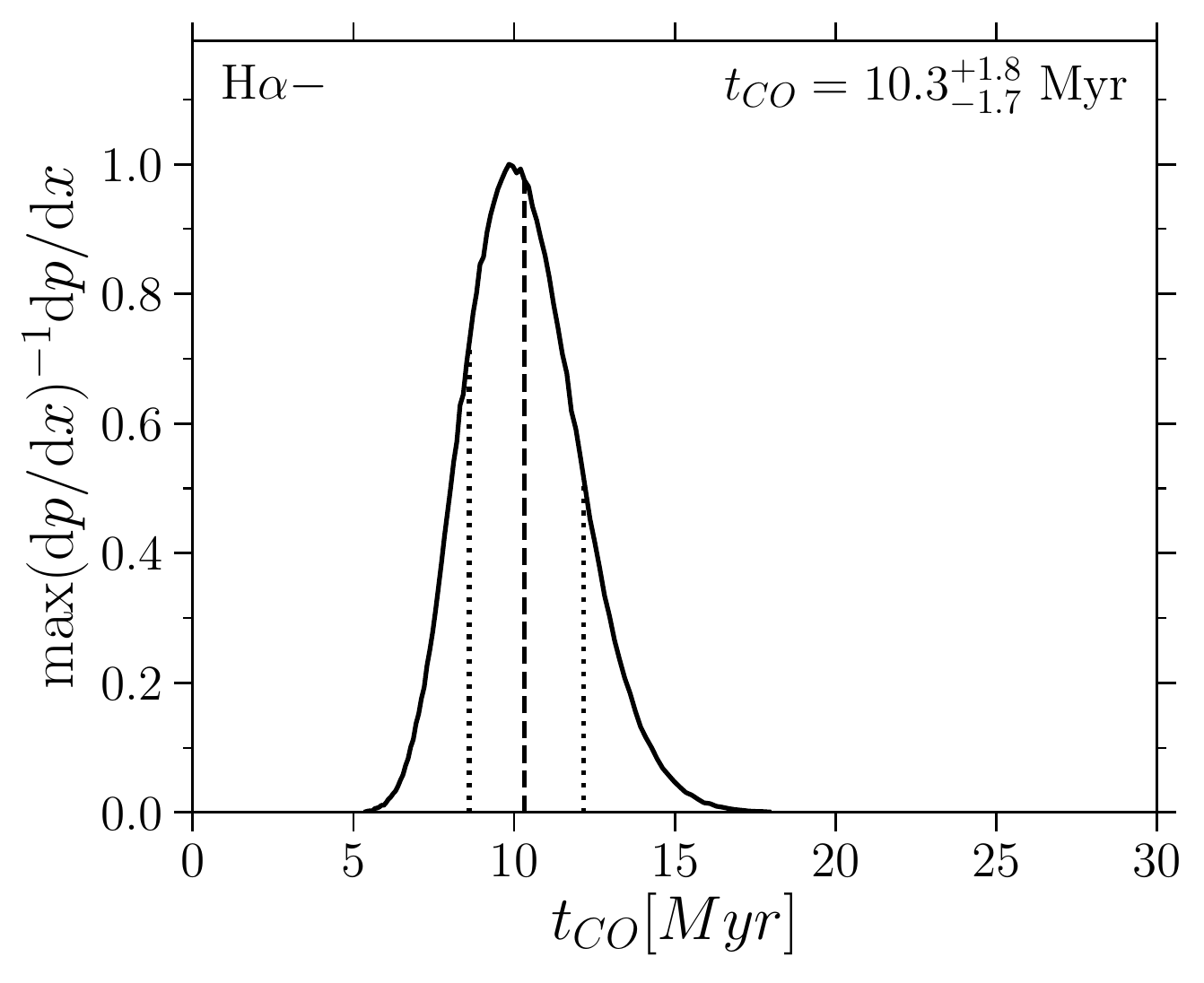}
	\includegraphics[width=0.49\linewidth]{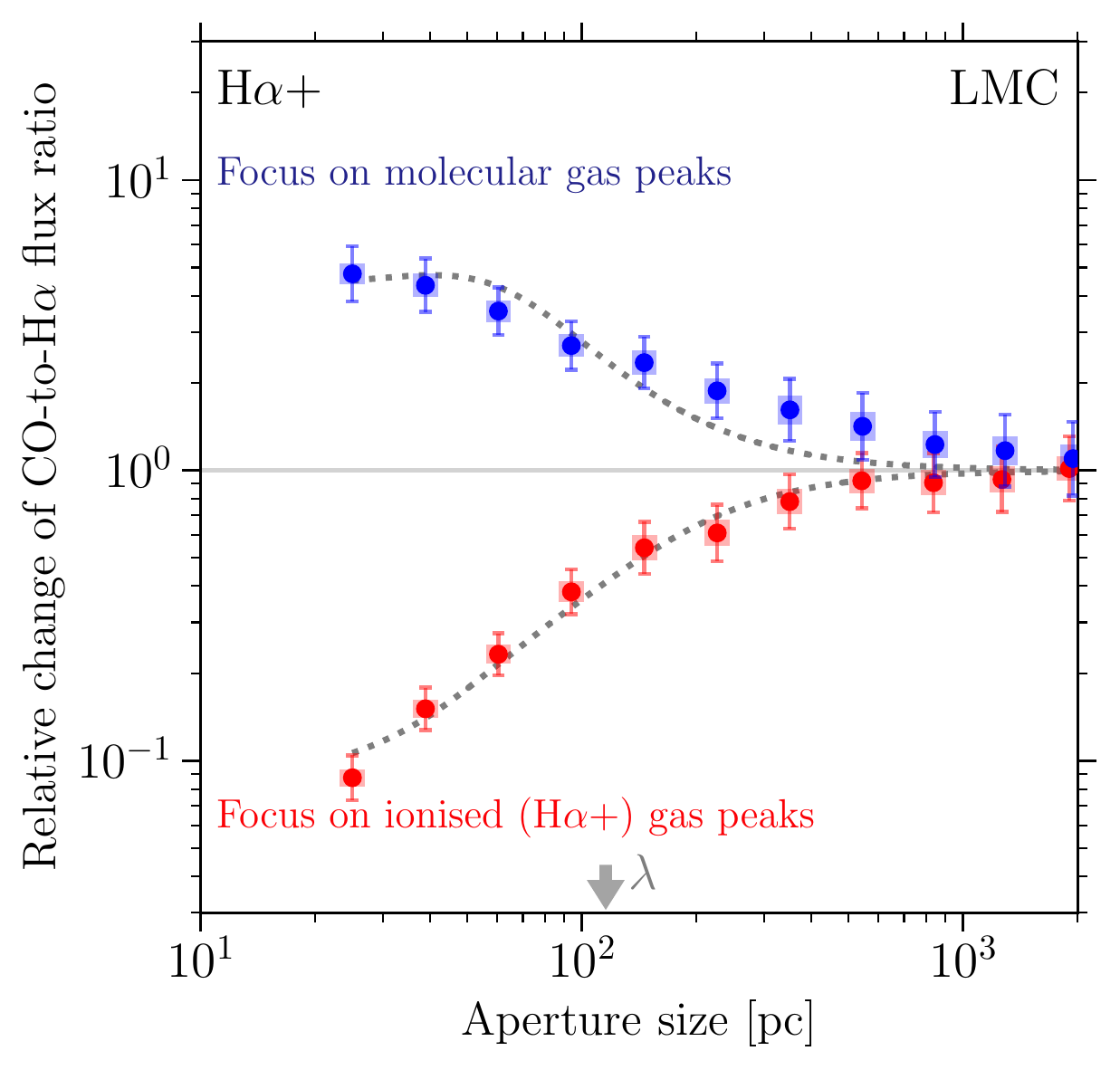}
	\includegraphics[width=0.49\linewidth]{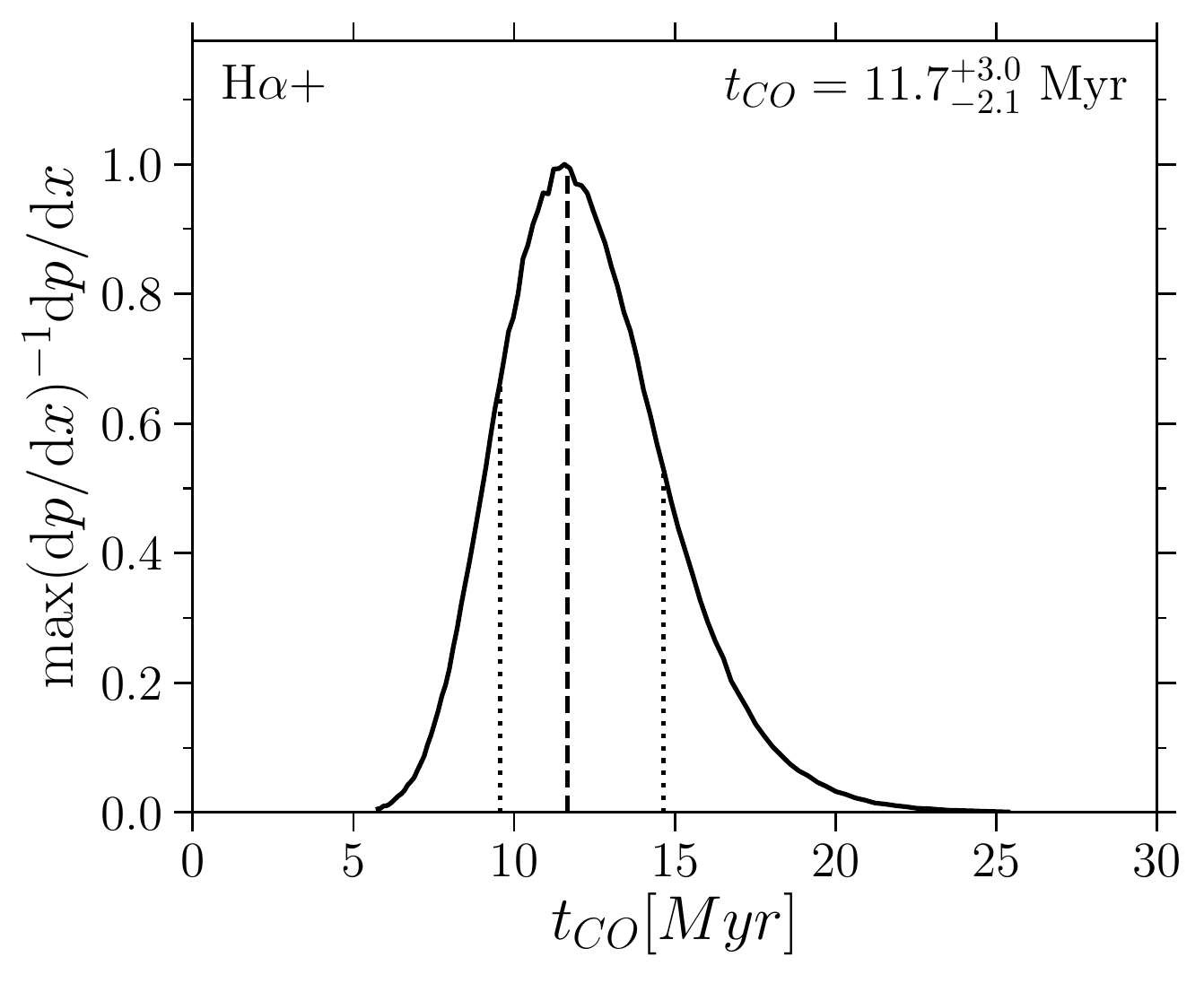}
	\caption{\label{nomasks}Repeat of our main analysis when \textit{not} masking the regions in which molecular clouds are detected by the NANTEN CO survey that are not covered by the MAGMA survey. Left: The gas-to-stellar flux ratio as a function of spatial scale. Right: one-dimensional PDFs of the inferred molecular cloud lifetime. Upper panels: The SHASSA continuum-subtracted H$\alpha$ image is used as the reference map. Lower panels: The MCELS H$\alpha$ image including continuum is used as the reference map. The results are entirely consistent with our fiducial results (see Section~\ref{results}). This indicates that our approach of masking the regions that contain molecular clouds but are not included by the MAGMA CO survey of the LMC does not bias our measurement of the molecular cloud lifetime.}
\end{figure*}

Because the MAGMA survey strategy focused primarily on the brightest CO clouds from the NANTEN survey and neglected the faintest sources, using the MAGMA survey data without masking regions of unobserved molecular clouds from the H$\alpha$ emission map (as is done in the main body of this work) is effectively the same as introducing a sensitivity limit to the survey, which is then used to mask the H$\alpha$ maps as well. To assess the impact of this choice, we repeat the experiment without masking the regions surrounding the clouds that are only detected in the NANTEN survey. The resulting CO-to-H$\alpha$ flux ratios and $t_{\text{CO}}$ PDF are presented in Figure~\ref{nomasks}, which shows that our results are insensitive to the masking strategy.

\begin{figure*}
	\includegraphics[width=0.49\linewidth]{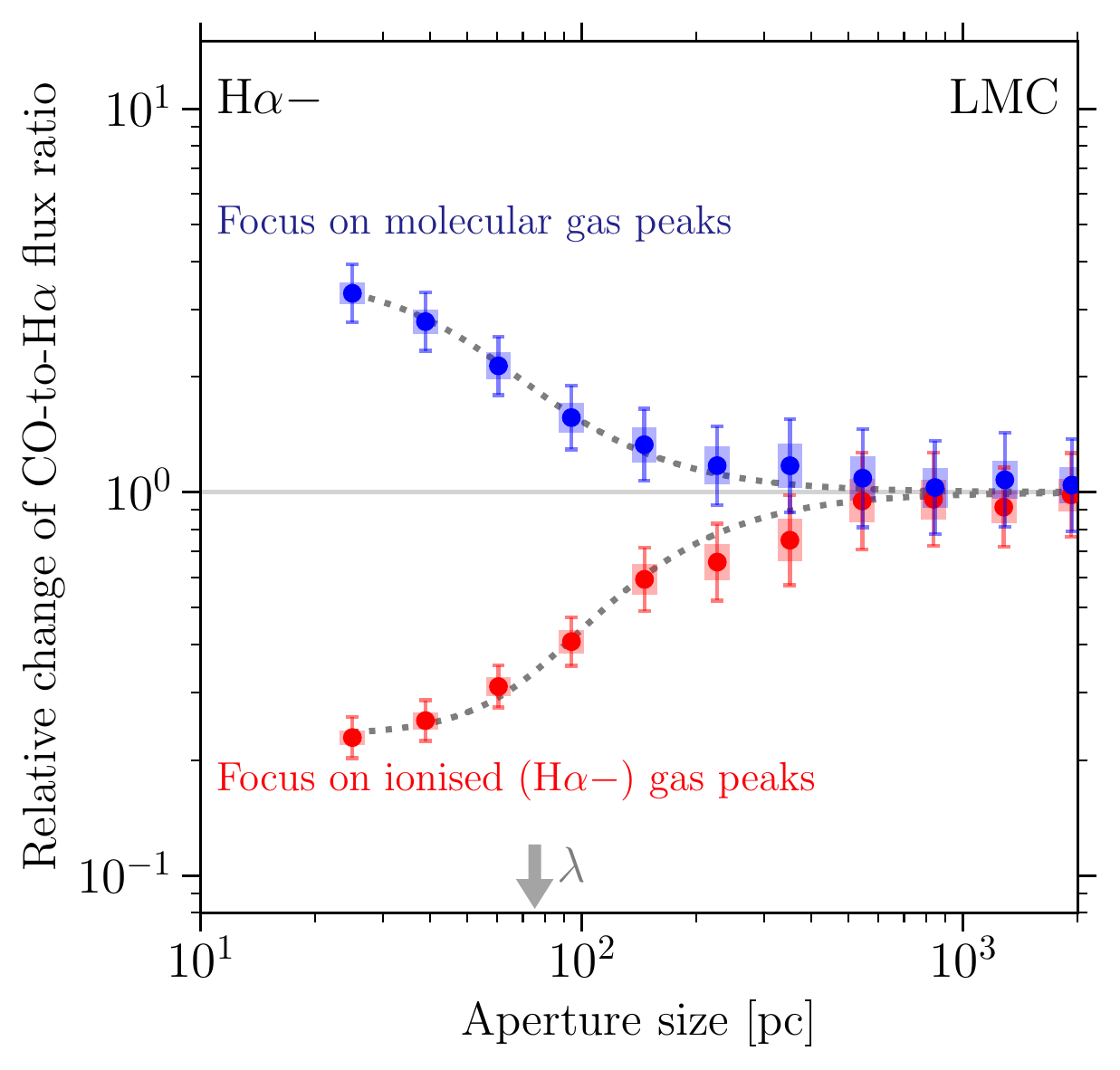}
	\includegraphics[width=0.49\linewidth]{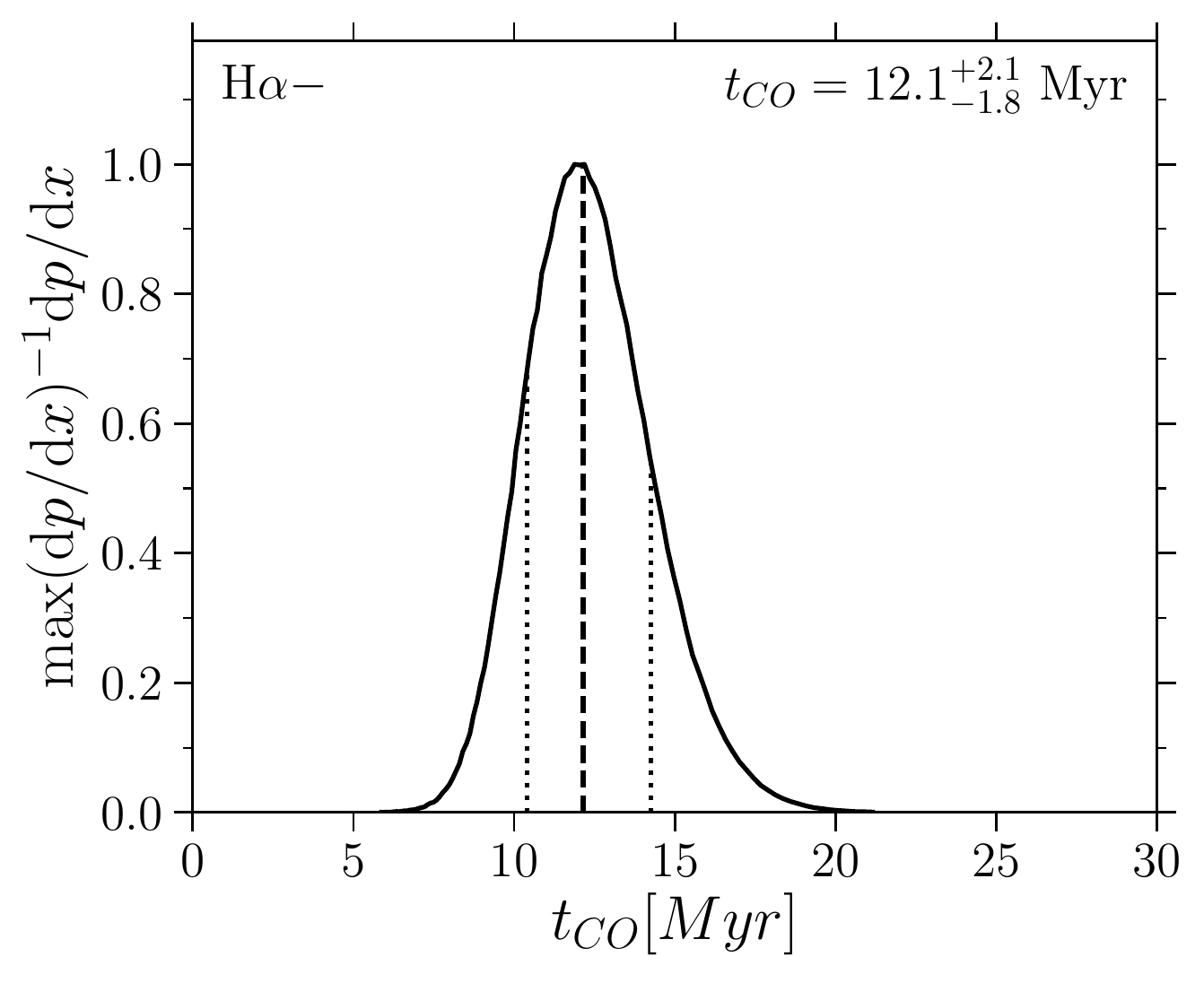}
	\includegraphics[width=0.49\linewidth]{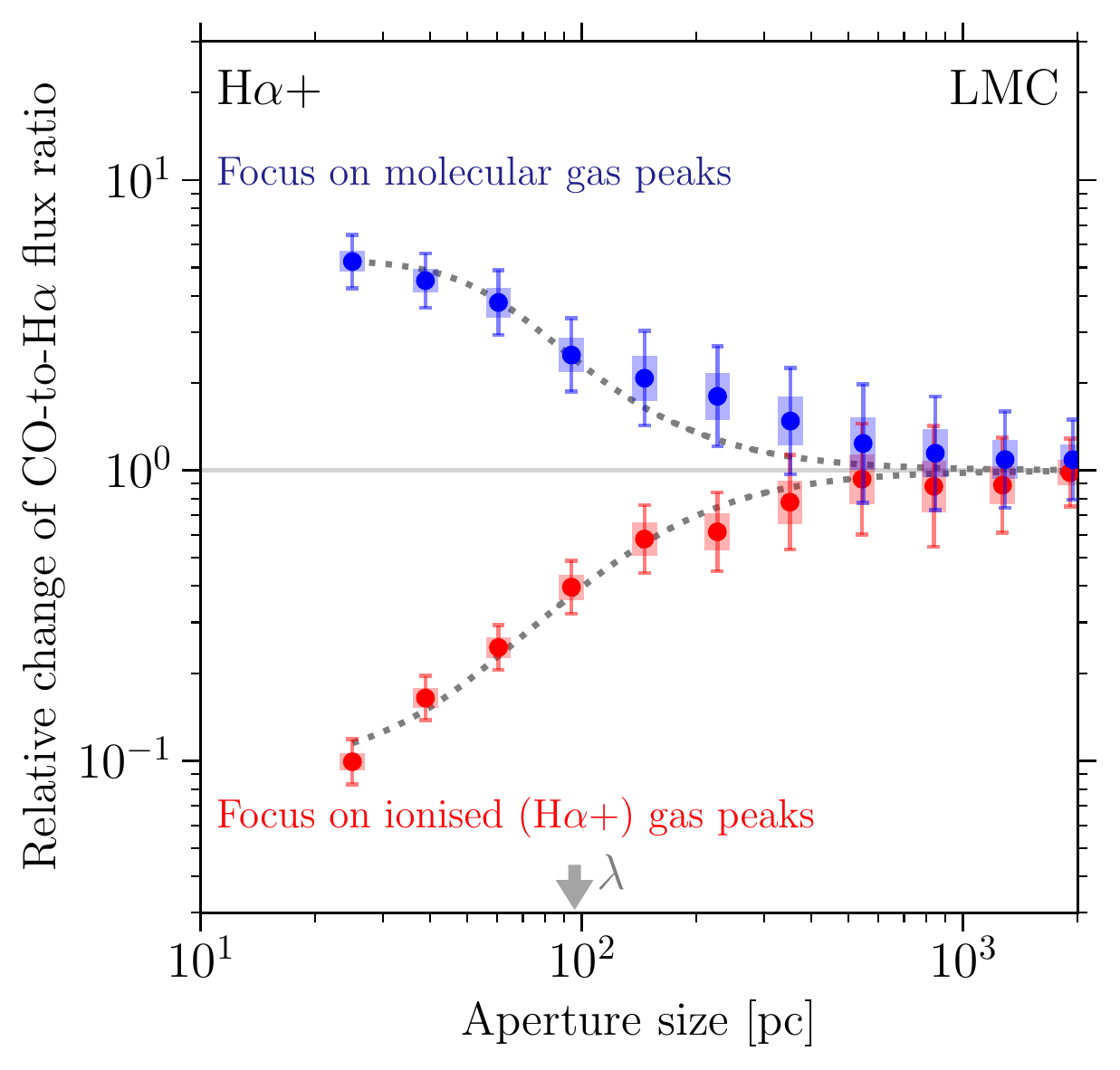}
	\includegraphics[width=0.49\linewidth]{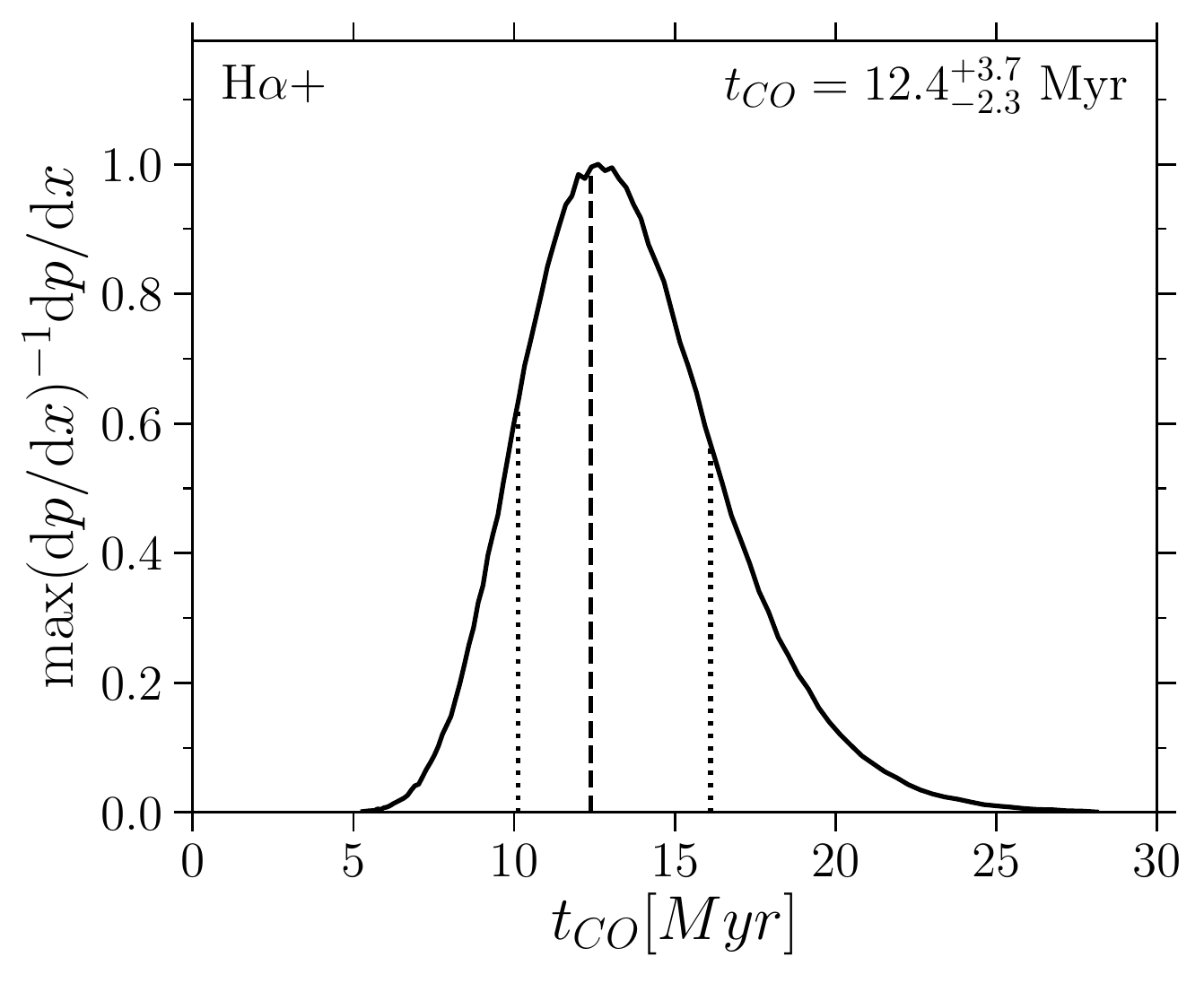}
	\caption{\label{combinedwithModels}Repeat of our main analysis when artificially adding model CO emission for molecular clouds detected by the NANTEN CO survey that are not covered by the MAGMA survey. Left: The gas-to-stellar flux ratio as a function of spatial scale. Right: one-dimensional PDFs of the inferred molecular cloud lifetime. Upper panels: The SHASSA continuum-subtracted H$\alpha$ image is used as the reference map. Lower panels: The MCELS H$\alpha$ image including continuum is used as the reference map. The results are entirely consistent with the time-scales derived without the addition of the model clouds (see Section~\ref{results}). This indicates that our fiducial approach of masking the regions that contain molecular clouds but are not included by the MAGMA CO survey of the LMC does not bias our measurement of the molecular cloud lifetime.}
\end{figure*}

We also test how the results are affected by the limited coverage of the MAGMA survey by adding the NANTEN-only detections as artificial CO-bright clouds represented by two-dimensional Gaussians, using their positions, sizes, and fluxes as reported in the NANTEN molecular cloud catalogue. The resulting CO-to-H$\alpha$ flux ratios and $t_{\text{CO}}$ PDF are presented in Figure~\ref{combinedwithModels}, which shows that our results are unaffected by the limited areal coverage of the MAGMA survey.

The constrained best-fitting parameters for both experiments carried out in this appendix are listed in Table~\ref{appendixTable}, along with the results from the two fiducial runs from Section~\ref{results} for clarity. Also shown are the goodness of fit parameters and the number of peaks selected in each tracer for each run. All constrained parameters are consistent with each other to within the uncertainties, even though the goodness of fit is lower for the experiment where the missing clouds from the MAGMA survey are modelled. Taken together, these results demonstrate the robustness of our approach. 

\begin{table*}
	\caption{\label{appendixTable}}
	\begin{tabular}{c c c c c c c c c c}
		\hline
		reference map & target map & NANTEN only clouds &  $n_{\text{ref}}$ & $n_{\text{tar}}$ & $t_{\text{ref}}$ & $t_{\text{CO}}$ & $t_{\text{fb}}$ & $\lambda$ & $\chi^{2}$ \\
		\hline
		H$\alpha$- & MAGMA CO & masked (fiducial) & 297 & 347 & 4.67$\substack{+0.15\\-0.34}$ & 11.4$\substack{+1.9\\-2.1}$ & 1.1$\substack{+0.2\\-0.2}$ & 79$\substack{+18\\-13}$ & 0.28 \\
		H$\alpha$- & MAGMA CO & unmasked & 307 & 343 & 4.67$\substack{+0.15\\-0.34}$ & 10.3$\substack{+1.8\\-1.7}$ & 1.1$\substack{+0.2\\-0.2}$ & 81$\substack{+18\\-11}$ & 0.39 \\
		H$\alpha$- & MAGMA CO & modelled & 324 & 403 & 4.67$\substack{+0.15\\-0.34}$ & 12.1$\substack{+2.1\\-1.8}$ & 1.1$\substack{+0.3\\-0.2}$ & 75$\substack{+14\\-10}$ & 0.36 \\
		H$\alpha$+ & MAGMA CO & masked (fiducial) & 276 & 342 & 8.54$\substack{+0.97\\-0.82}$ & 12.3$\substack{+3.1\\-2.3}$ & 1.4$\substack{+0.5\\-0.3}$ & 109$\substack{+22\\-13}$ & 2.10\\
		H$\alpha$+ & MAGMA CO & unmasked & 274 & 338 & 8.54$\substack{+0.97\\-0.82}$ & 11.7$\substack{+3.0\\-2.1}$ & 1.4$\substack{+0.4\\-0.3}$ & 115$\substack{+22\\-14}$ & 2.24 \\
		H$\alpha$+ & MAGMA CO & modelled & 281 & 403 & 8.54$\substack{+0.97\\-0.82}$ & 12.4$\substack{+3.7\\-2.3}$ & 1.1$\substack{+0.5\\-0.4}$ & 96$\substack{+23\\-18}$ & 0.80 \\
		\hline
	\end{tabular}
\end{table*}


\bsp	
\label{lastpage}
\end{document}